\documentclass[prl,twocolumn,preprintnumbers,amssymb,amsmath, 
superscriptaddress]{revtex4} 

\pdfoutput=1

\usepackage{graphicx}
\usepackage[caption=false]{subfig}
\usepackage{bm}
\usepackage{color}
\usepackage{times}
\usepackage[colorlinks,bookmarks=false,citecolor=blue,linkcolor=red,urlcolor=blue]{hyperref}
\newcommand{\be}{\begin{equation}}
\newcommand{\ee}{\end{equation}}
\newcommand{\bea}{\begin{eqnarray}}
\newcommand{\eea}{\end{eqnarray}}

\newcommand\blue[1]{{\color{blue}{#1}}}

\begin{document}
%%%%%%%%%%%%%%%%%%%%%%%%%%%%%%%%%%%%%%%%%%%%%%%%%%%%%%%%%%%%%%%%%%%%%%%%%%%%%%
\title{Real time confinement following a quantum quench to a non-integrable model}
%%%%%%%%%%%%%%%%%%%%%%%%%%%%%%%%%%%%%%%%%%%%%%%%%%%%%%%%%%%%%%%%%%%%%%%%%%%%%%
\author{M. Kormos}
\affiliation{MTA-BME "Momentum" Statistical Field Theory Research Group, 1111 Budapest, Budafoki {\'u}t 8, Hungary}
\author{M. Collura}
\affiliation{SISSA and INFN, via Bonomea 265, 34136 Trieste, Italy. }
\author{G. Tak{\'a}cs}
\affiliation{MTA-BME "Momentum" Statistical Field Theory Research Group, 1111 Budapest, Budafoki {\'u}t 8, Hungary}
\affiliation{Department of Theoretical Physics, Budapest University of Technology and Economics 1111 Budapest, 
Budafoki {\'u}t 8, Hungary}
\author{P. Calabrese}
\affiliation{SISSA and INFN, via Bonomea 265, 34136 Trieste, Italy. }
\begin{abstract}

Light cone spreading of correlations and entanglement is a key feature of the non-equilibrium quench dynamics 
of many-body quantum systems.
First proposed theoretically \cite{cc-06}, it has been experimentally revealed in cold-atomic gases \cite{cetal-12,kauf} 
and it is expected to be a generic characteristic of any quench in systems with short-range interactions and no disorder. 
Conversely, here we propose a mechanism that, through confinement of the elementary excitations, 
strongly suppresses the light-cone spreading. 
Confinement is a celebrated concept in particle physics, but it also exists in condensed matter systems, 
most notably in one spatial dimension where it has been experimentally observed \cite{exp}.
Our results are obtained for the Ising spin chain with transverse and longitudinal magnetic field, but the proposed mechanism is of general validity since it is based on the sole concept of confinement and it should be easily observed in cold atom experiments.

\end{abstract}
%\pacs{64.70.Tg, 03.67.Mn, 75.10.Pq, 05.70.Jk}
\maketitle

It is widely known that some  fundamental constituents of matter, the quarks, cannot be observed free in nature because they are confined into baryons and mesons, as a consequence  of the fact that the strong interaction between them increases with their separation. 
It is less known that this phenomenon also occurs in condensed matter and statistical physics 
as nowadays experimentally demonstrated in several quasi one-dimensional compounds \cite{exp}. 
Most of the theoretical and experimental studies so far concentrated on understanding the consequences of confinement for the equilibrium physics of both high energy and condensed matter systems. 
Here instead we show that confinement has dramatic consequences for the non-equilibrium dynamics following a quantum quench and that this can also be used effectively as a quantitative probe of confinement. 

A global quantum quench is the non-equilibrium dynamics initiated by a sudden change of a parameter in the Hamiltonian of an isolated quantum system, a protocol which is routinely engineered in cold-atom 
experiments \cite{cetal-12,kww-06,tc-07,tetal-11,getal-11,mmk-13,fse-13,lgk-13,kauf}.
According to a by now standard picture \cite{cc-06}, the initial state acts as a source of quasiparticles excitations.
A quasiparticle  of momentum $p$ moves with velocity $v_p$ and carries quantum correlations through the systems. 
Indeed, particles emitted from regions of size of the initial correlation length are entangled, while particles created far from each other
are incoherent. 
If there is a maximum speed of propagation $v_{\rm max}\geq v_p$ (e.g. as a consequence of the 
Lieb--Robinson bound \cite{lr-72}) all connected correlations at distance $\ell$ vanish for  
times such that $2v_{\rm max} t< \ell$ \cite{cc-06} and the entanglement entropy of an interval of length $\ell$ 
grows linearly in the same time window \cite{cc-05}.
This scenario has been confirmed by numerous exact calculations and numerical simulations (see e.g. \cite{lc-sol}) both in integrable and nonintegrable models and tested in real experiments both for correlations \cite{cetal-12} 
and entanglement \cite{kauf}. 
To the best of our knowledge, up to now violations of light cone spreading have only been observed in models with quenched disorder \cite{qd} and systems with long-range interactions \cite{lr}, but these lie beyond the range of applicability of the above argument.

 \begin{figure}[b]
 \includegraphics[width=.35\textwidth]{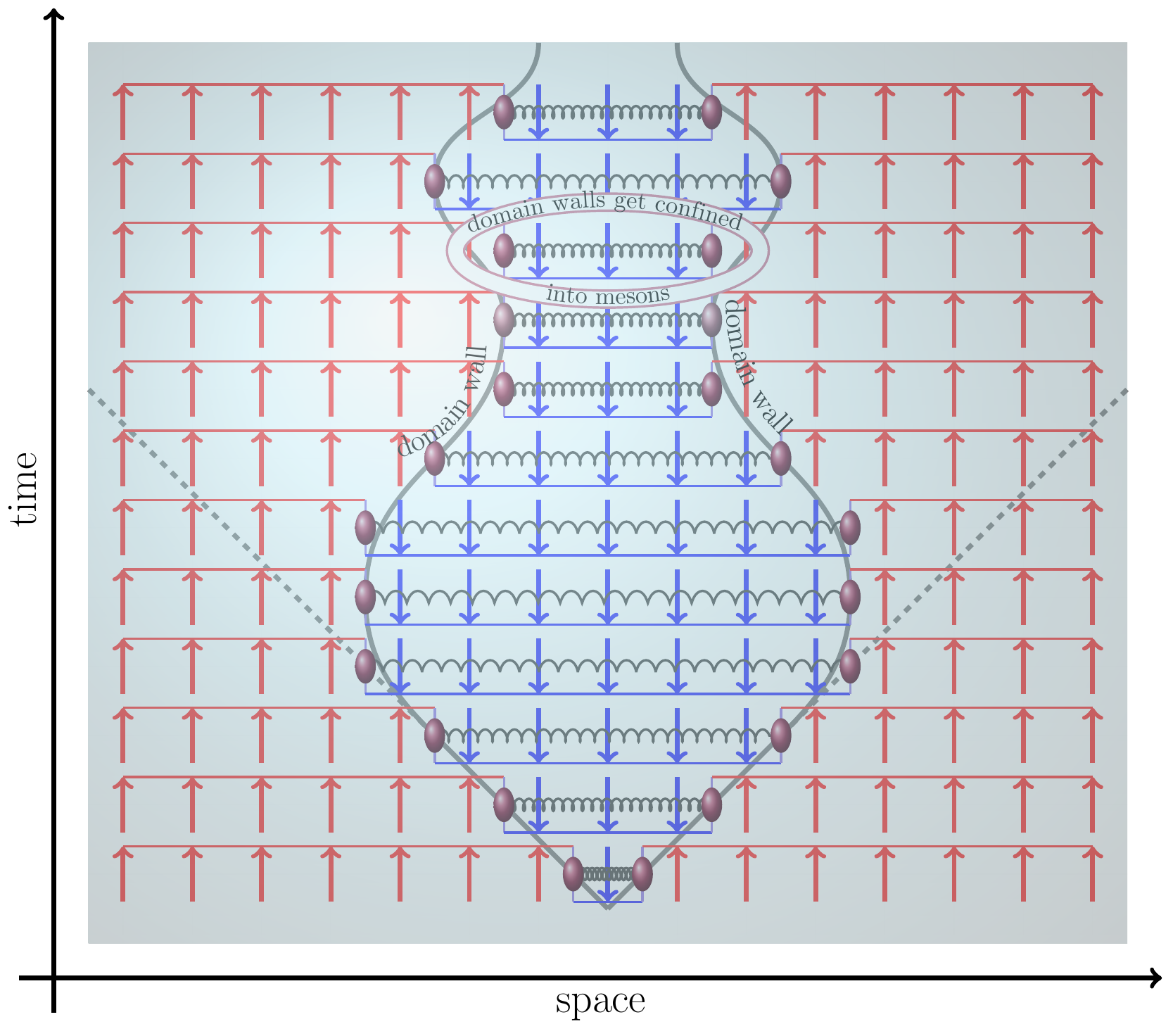}
 \caption{Pictorial semiclassical picture of a meson state in the Ising model:
 two counter-propagating domain walls %(i.e. a domain with magnetisation opposite to the field $h_x$) 
 bounce back and forth because of a confining interaction.  
 }
 \label{fig:sketch}
 \end{figure}

How can confinement change qualitatively the spreading of correlations?
Exactly as in the standard scenario, the initial state acts as a source of quasiparticles. 
Pairs of quasiparticles move in opposite directions, but due to the confining potential the farther they go apart
the stronger is the attractive force they feel, which eventually turns the particles back leading to an oscillatory behaviour,
as depicted in Fig. \ref{fig:sketch}. 
In analogy to strong interaction physics, the resulting bound states are called mesons. 
After a generic quench, mesons can be produced in pairs of opposite momenta, but being (as we shall see) 
heavy massive particles they move much slower than the elementary quasiparticles, 
resulting in light-cone phenomena which are qualitatively different from the unconfined case. 
 
Confinement is known to take place in one of the paradigmatic models of statistical mechanics, namely  
the Ising chain in both  transverse ($h_z$) and longitudinal ($h_x$) magnetic fields with Hamiltonian 
\be
H=-J \sum_{j=-\infty}^{\infty}\left[\sigma_j^x\sigma_{j+1}^x + h_z\sigma_j^z + h_x\sigma_j^x\right]\,,
\label{ham}
\ee
where $\sigma_j^\alpha$ are the Pauli matrices.
This model has been already engineered with cold atoms \cite{coldising} and quench protocols have been  
 implemented by using Feshbach resonance \cite{mmk-13}.

For $h_x=0$, the Hamiltonian (\ref{ham}) can be diagonalised by a Jordan--Wigner mapping to free 
spinless Majorana fermions with the dispersion relation $\epsilon(k) = 2J \sqrt{1-2h_z \cos k +h_z^2}$ \cite{SM}.
At $h_z=1$ the system has a quantum critical point separating the paramagnetic and ferromagnetic phases. 
For $h_z<1$ the system is in the gapped ferromagnetic phase where the massive fermions can be thought of as freely 
propagating domain walls separating domains of magnetisation $\bar\sigma=(1-h_z^2)^{1/8}.$  
Switching on a non-zero field $h_x$ induces a linear attractive potential between pairs of domain walls which enclose a 
domain of length $d$ and of magnetisation opposite to $h_x$. 
For small $h_x$, the potential can be approximated as $V(d)=\chi\cdot d$ with $\chi=2Jh_x\bar\sigma$ \cite{mw-78}. 
As a result, the domain walls are confined into bound states (mesons), 
a scenario first proposed by McCoy and Wu \cite{mw-78}. 
For $h_z>1$ and $h_x\neq0$ the physics is very different and there is no confinement \cite{zam-13}.
The model for $h_x\neq0$ is no longer integrable and {\blue the spectrum can only be described by resorting to 
various approximations, such as e.g. field-theoretical ones \cite{zam-13,m-vari,ts-vari} that are valid in 
the vicinity of the critical point $h_z=1$}. 
Here we use a different approach: the low-density approximation of Ref. \cite{r-08}, which describes the energy levels very accurately 
when the system is far away from the critical point. In the Supplementary Material \cite{SM} we report the details of this approximation and several pieces of evidence of its applicability for the values of magnetic fields of interest. 
This approximation allows us to calculate all the properties of the mesons we need, namely their number, 
masses, and velocities.

To simulate the time evolution of the quantum quench we use an iTEBD \cite{iTEBD} algorithm, 
details of which are reported in \cite{SM}. 
In the following we report numerical results for several observables, building up the evidence 
that the observed dynamics is governed by confinement effects and it is dramatically different 
from the unconfined one.

\begin{figure}[t]
 \includegraphics[width=.22\textwidth]{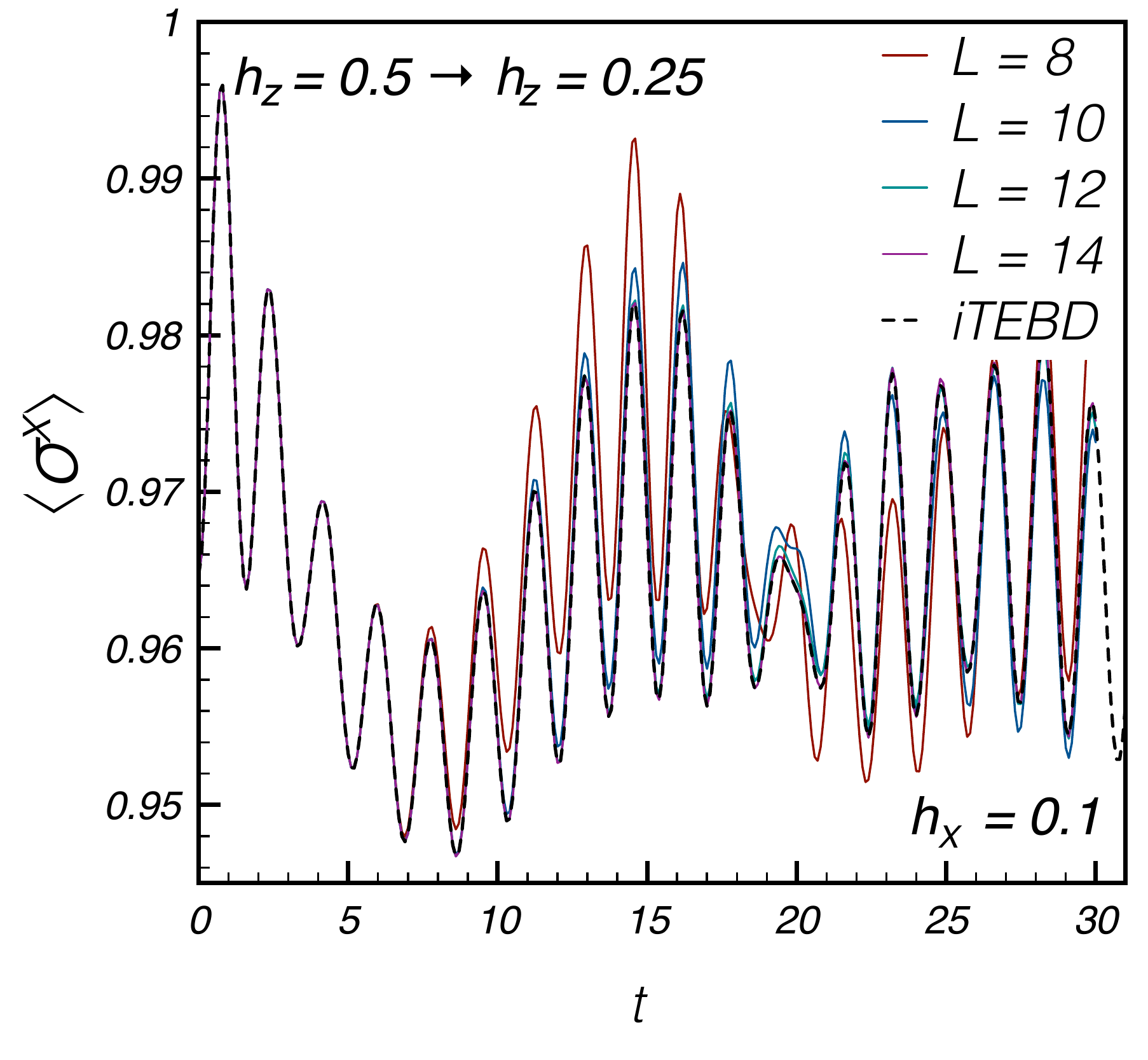} 
\includegraphics[width=.22\textwidth]{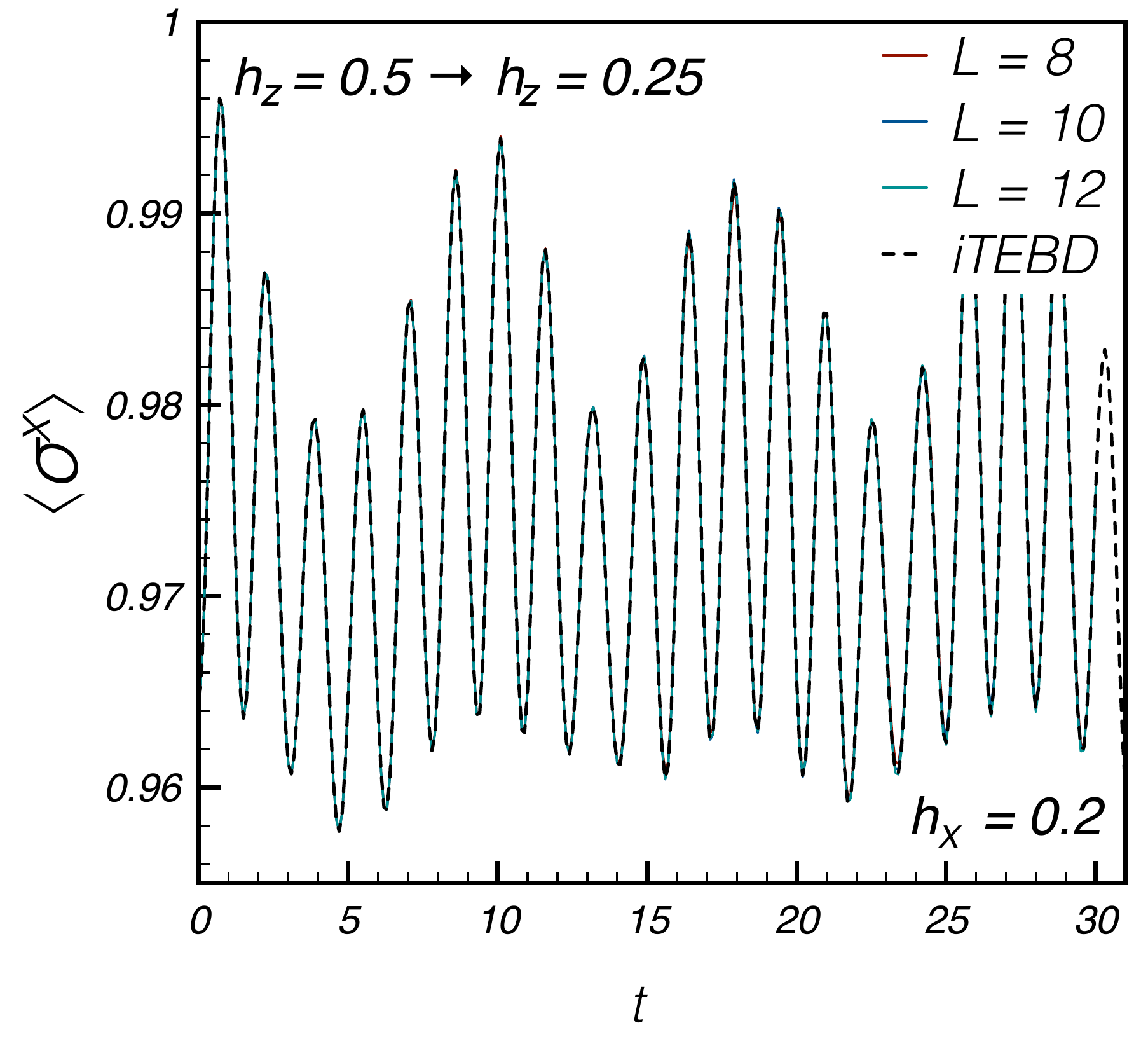}\\
\includegraphics[width=.22\textwidth]{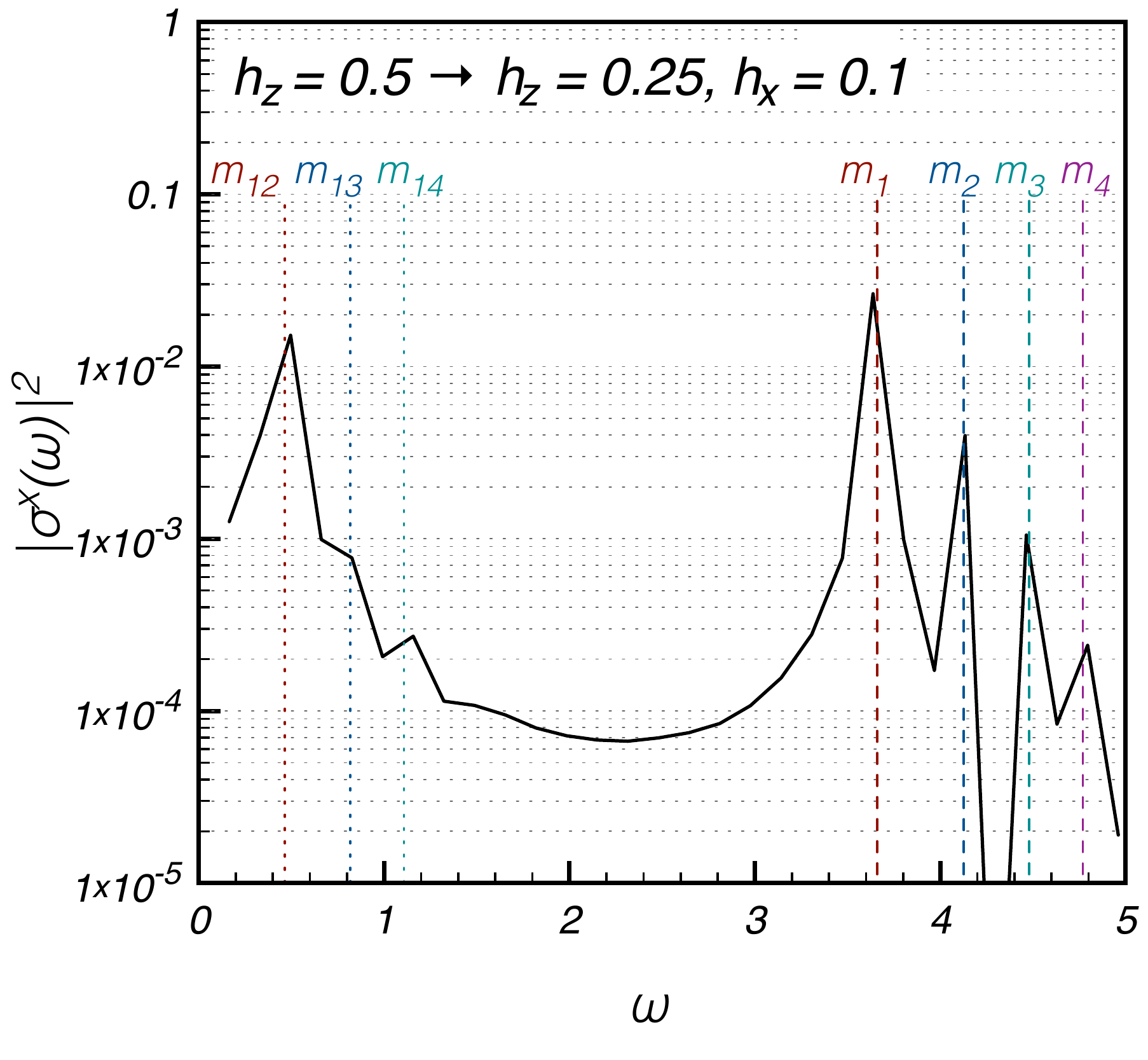}
\includegraphics[width=.22\textwidth]{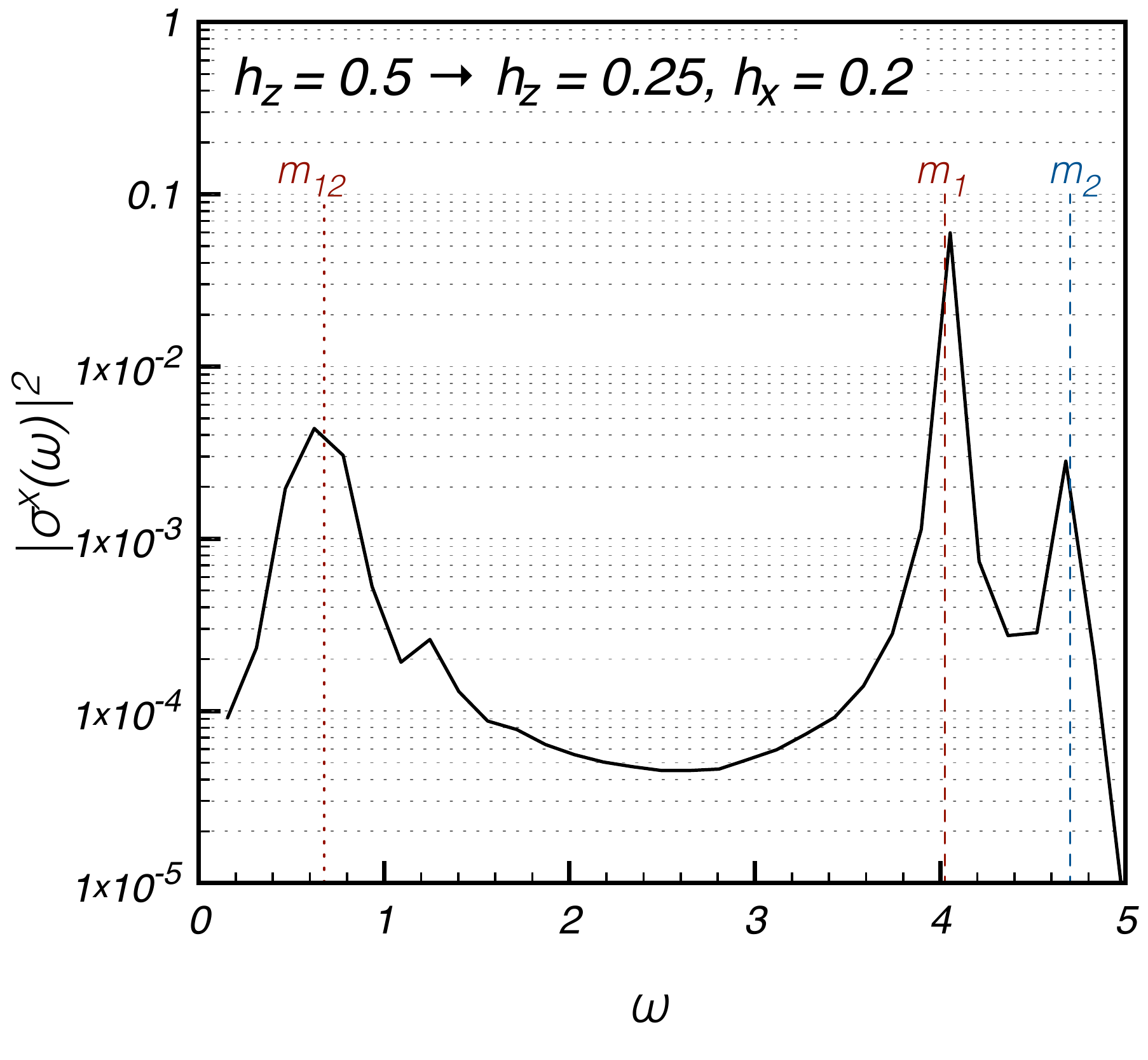}
  \caption{ {\it Upper panels}: 
Time evolution of the longitudinal magnetisation $\langle{\sigma^x(t)}\rangle$ after quenching from $h_{z}=0.5,h_{x}=0$ 
to $h_z=0.25$ and  $h_x=0.1, 0.2$. 
Dots are iTEBD results, lines are exact diagonalisation results for $L=8,\dots,12,$.
% dashed lines are diagonal ensemble expectation values which are expected to be reached asymptotically. 
{\it Lower panels}: power spectrum of $\langle{\sigma^x(t)}\rangle$ in which the dashed vertical lines show the  
meson masses and their differences.}
\label{fig:sigmat}
\end{figure}

{\it Expectation value of the order parameter}.  
We first consider the time evolution of the order parameter, i.e. the  magnetisation  $\langle{\sigma^x(t)}\rangle$. 
We recall that in the integrable case with vanishing longitudinal field, 
$\langle{\sigma^x(t)}\rangle$ decays to zero exponentially for any quench within the ferromagnetic phase \cite{cef-11} (see also \cite{SM}). 
For non-zero $h_x$, we report the iTEBD data for $\langle{\sigma^x(t)}\rangle$ in Fig. \ref{fig:sigmat} (top) 
for two representative quenches, 
but the results are qualitatively the same for all the values of initial and final fields we considered within the ferromagnetic phase. 
It is evident from the figure that a small longitudinal field radically alters the dynamics, turning the exponential 
relaxation into an oscillatory behaviour with numerous different frequencies. 
The qualitative change of the dynamics is the consequence of confinement. 
This can be demonstrated by extracting the oscillation frequencies with a discrete Fourier transform of the time series,  
which are reported in Fig. \ref{fig:sigmat} (bottom). 
The dominant frequencies in the resulting power spectrum are compatible, to a surprising high degree of accuracy, 
with the masses of the mesons  and  their  differences (obtained explicitly in \cite{SM}).

On the more technical side, we observe that this dynamics shows rather weak finite size effects.
Indeed, in the figure the iTEBD data (valid for infinite chains) are almost indistinguishable (up to a given time that grows with the chain length) from the exact diagonalisation results for chains of length between 8 and 12. 

{\it The two-point function} is the quantity that shows the strongest effects of confinement. 
Indeed, when the interaction $h_x$ is turned on, the propagating particles are the heavy massive mesons and not the 
light domain walls. 
Mesons propagate with a maximal velocity which is smaller than that of the domain walls: 
as shown in the Supplementary Material, together with the masses of the mesons, their dispersion relations and 
velocities can also be readily obtained in the zero-density approximation.

\begin{figure}[t]
 \includegraphics[width=.53\textwidth]{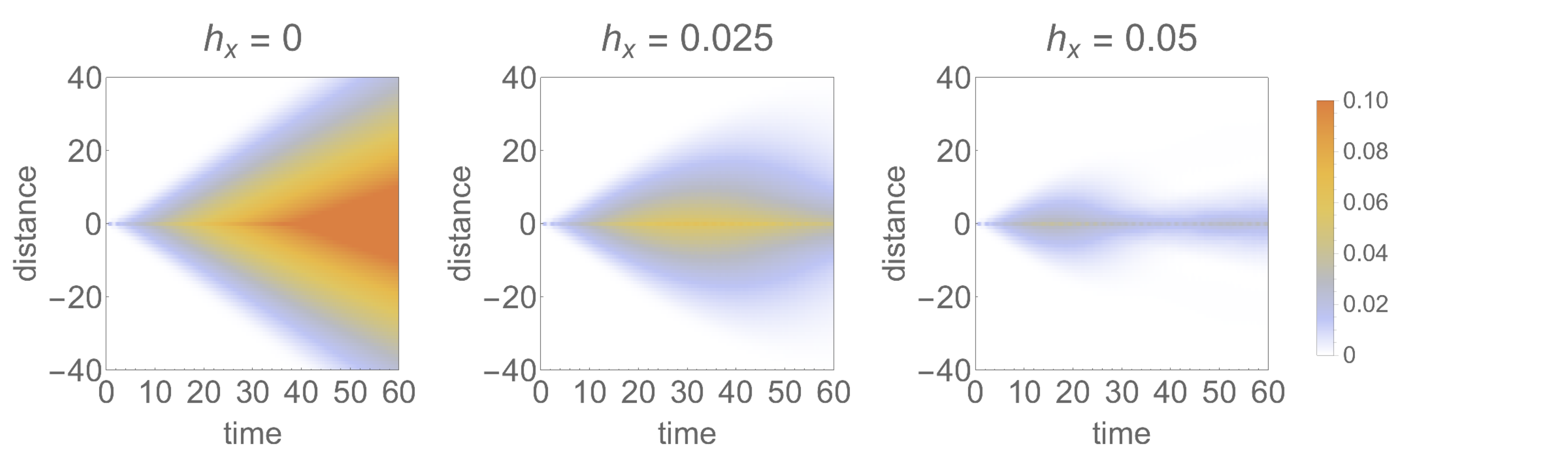}
 \includegraphics[width=.53\textwidth]{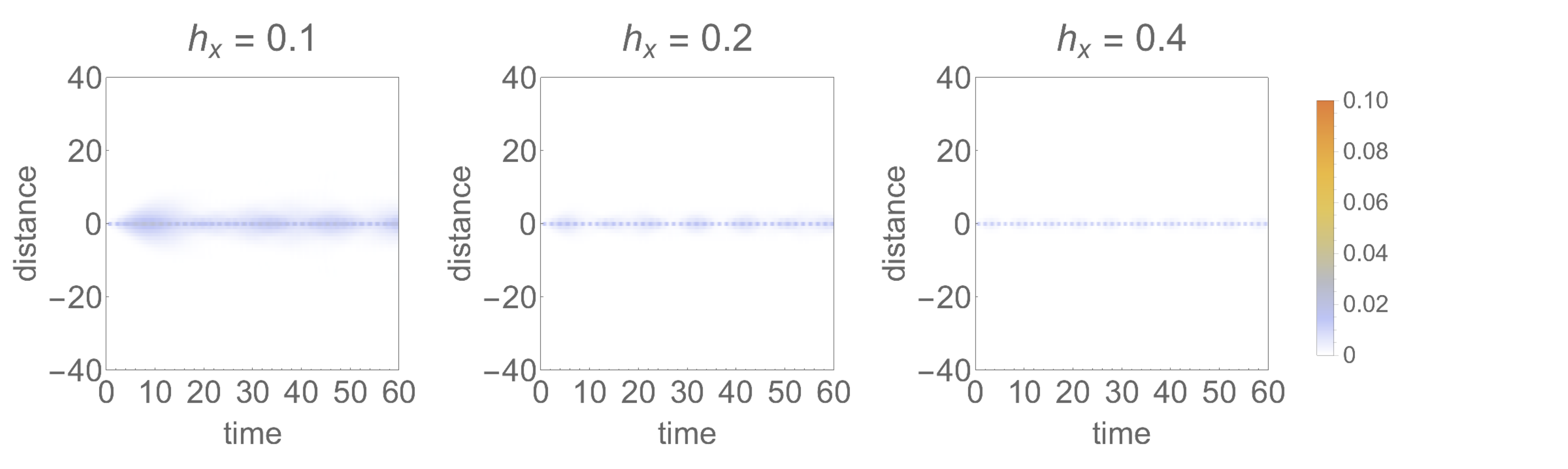}
  \caption{Connected longitudinal spin-spin correlation function $\langle{\sigma^x_1\sigma^x_{m+1}}\rangle_c$ 
  after quenching to the ferromagnetic point $h_z=0.25$ with a longitudinal magnetic field $h_x=0,0.025,0.05,0.1,0.2,0.4$.}
\label{fig:lss}
\end{figure}

However, it turns out that the effect of confinement in some cases is even stronger than an already dramatic and non-perturbative 
change of speed of propagation. 
Let us, for example, consider the quench from the fully ferromagnetic state (all spins up, i.e. the ground-state at $h_z=0$) 
to the points with $h_z=0.25$ and varying $h_x$ from 0 to $0.4$. 
In Fig. \ref{fig:lss} we report the equal time connected longitudinal spin-spin correlation function 
$\langle{\sigma^x_1\sigma^x_{m+1}}\rangle_c$. 
If $h_x=0$, we recover the integrable dynamics that was solved exactly in \cite{cef-11} with a clear light cone spreading.
For a small value of $h_x=0.025$, we see that for relatively short times (up to $t\sim 20=2/h_x$ in units of $J$)
the correlation follows qualitatively the integrable behaviour, but then it gets drastically slowed down and bounces back. 
By further increasing $h_x$, the region where there is light cone propagation shrinks to an almost invisible portion of the space-time. 
What happens is that due to the heavy masses of the mesons, the quench only provides sufficient energy
to produce them at rest. This can be understood quantitatively for small $h_x$. 
For a quench with $h_x=0$, the initial state can be written in term of the post-quench eigenstates as \cite{cef-11}
\be
|\psi_0\rangle= \prod_{k>0} (1+ i K(k) a^\dag_k a^\dag_{-k}) |0\rangle,
\label{IniS}
\ee
where the explicit form of $K(k)$ is given in  \cite{SM}.
Let us add now a field $h_x$ in the post-quench Hamiltonian that confines the fermions into mesons. 
If $K(k)$ is small, the state is dominated by the linear terms in the expansion of the above product and 
there are only pairs of fermions with momentum $(k,-k)$. 
These can only form mesons of {\it zero momentum}.
Quadratic terms like $K(k) K(k') |k,-k,k',-k'\rangle $ are small corrections but they produce 
pairs of mesons with non-zero momenta, e.g. $\pm (k \pm k')$ (higher order terms give rise to similar pairs). 
When instead $K(k)$ is of order one, all these terms are of the same order and mesons propagate (in pairs) with their own velocity after the quench.
 
\begin{figure}[t]
 \includegraphics[width=.53\textwidth]{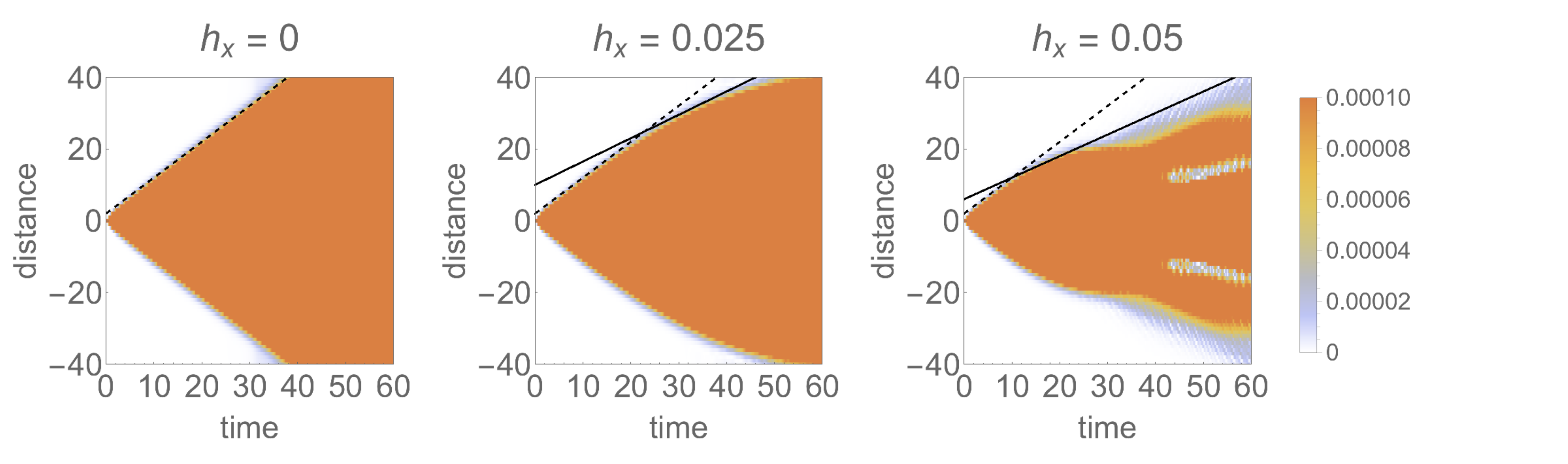}
 \includegraphics[width=.53\textwidth]{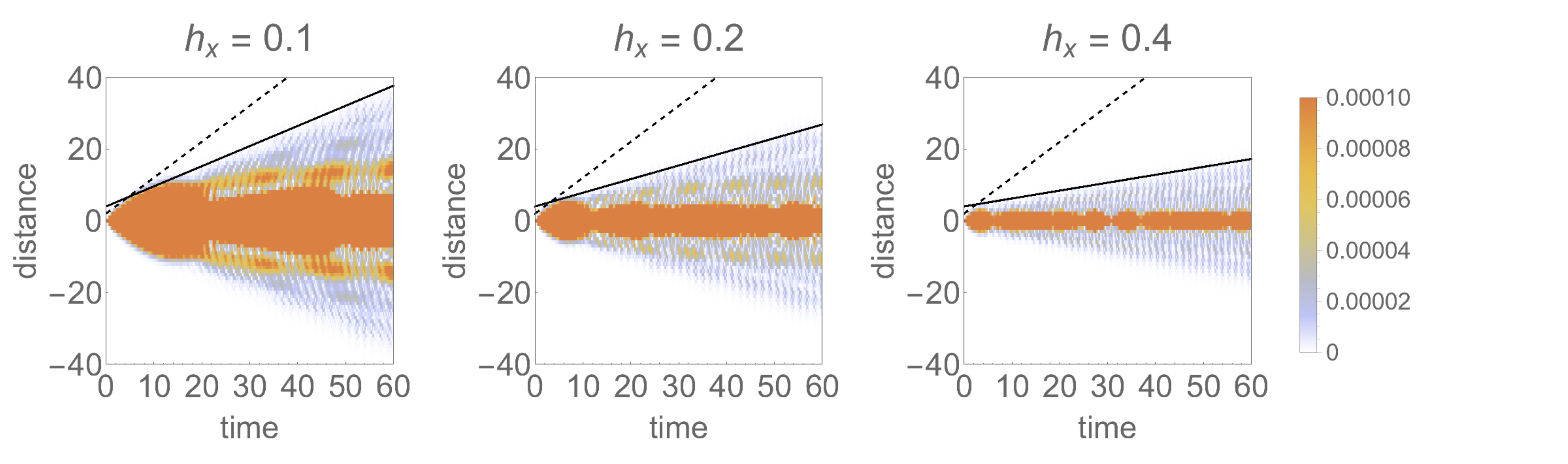}
  \caption{The same correlation function in Fig. \ref{fig:lss} but on a different scale: 
  the plotted signal is around $10^{-3}$ times the one in Fig. \ref{fig:lss} and the orange regions represent out of range values
   of the correlation function.}
\label{fig:lssa}
\end{figure}

For the quench reported in Fig. \ref{fig:lss} (from $h_z=0$ to $h_z=0.25$), $K(k)$ is small ($\ll1$) for all 
momenta and practically only zero-momentum mesons are formed. 
It should be however clear that zooming in the "white" region (i.e. the one apparently without signal) 
in Fig. \ref{fig:lss}, traces of mesons with non-zero velocities should be visible. 
For this reason, we report in Fig. \ref{fig:lssa} the same connected correlation already displayed in Fig. \ref{fig:lss}
but on a different intensity scale. 
Here, the orange regions correspond to values that are out of range.
All the visible signal of Fig. \ref{fig:lss} falls in these regions. 
The signal displayed in  Fig. \ref{fig:lssa} is approximately 3 orders of magnitude smaller than that in Fig. \ref{fig:lss} 
and shows a feeble light cone characterised by a velocity different from that of the domain walls (dashed lines).
We report the maximum value of the meson velocity (full lines) obtained in the low density approximation: 
it is evident that for all values of $h_x$ this velocity describes incredibly well the 
slope of the light cone. 

It is very important to stress once more that these confinement effects are non-perturbative:
a very small perturbation such as $h_x=0.025$ is enough to destroy completely the sharp light cone of the integrable model. 
%to a level that cannot be observed anymore in any experiment. 

\begin{figure}[t]
 \includegraphics[width=.54\textwidth]{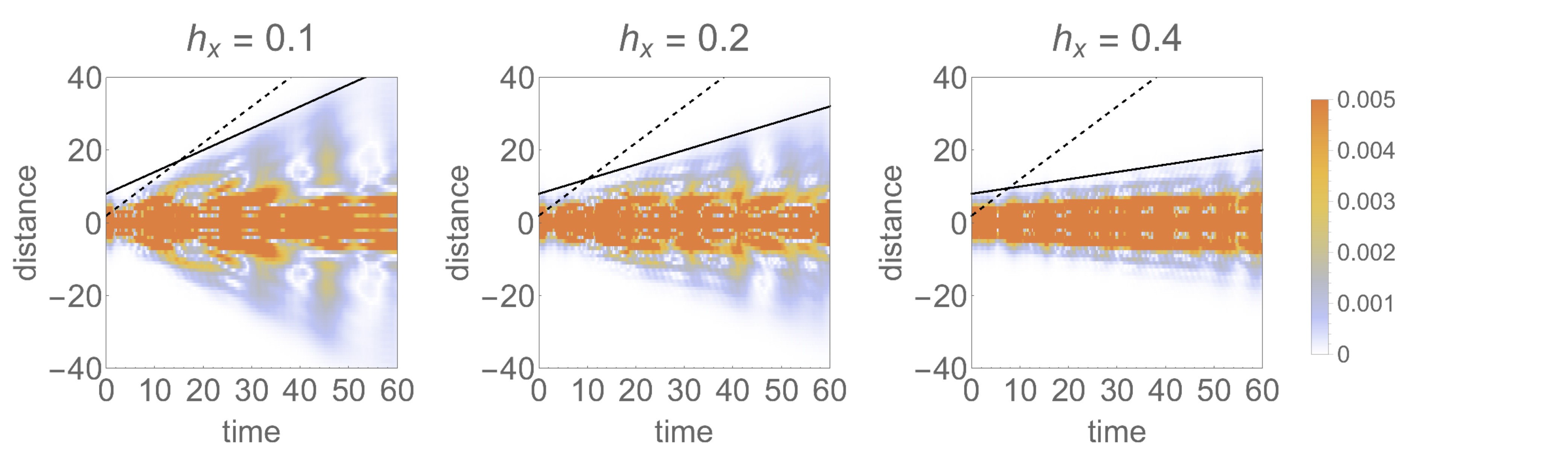}
 \includegraphics[width=.54\textwidth]{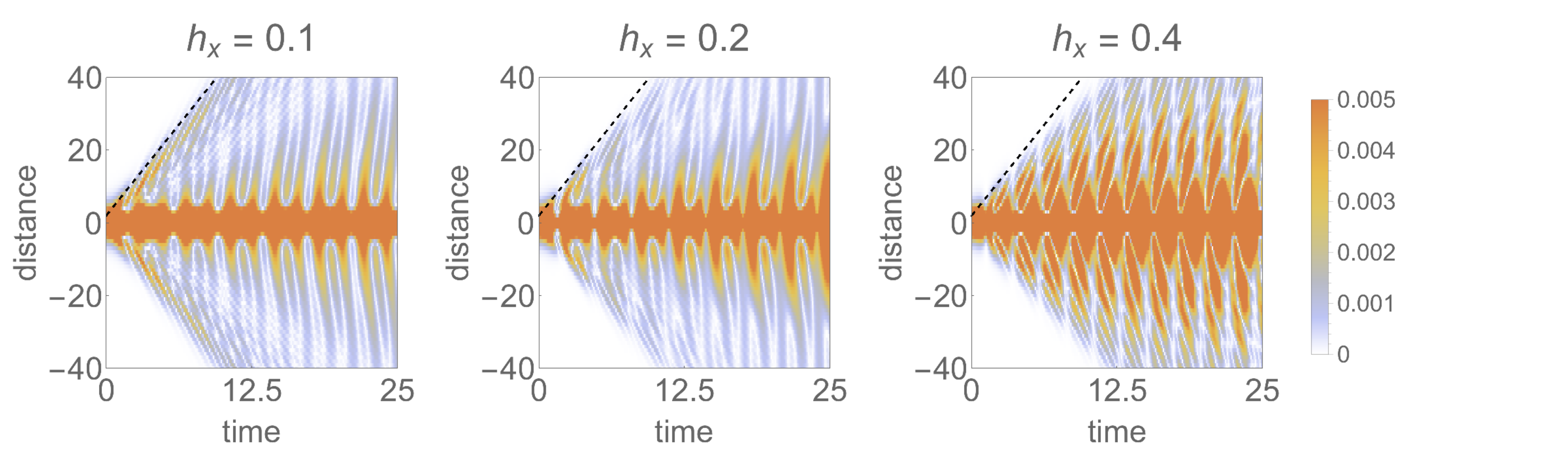}
 \caption{Absolute value of the correlation function $\langle{\sigma^x_1\sigma^x_{m+1}}\rangle_c$ for:
 (upper panel) quench from the paramagnetic phase ($h_z=2, h_x=0$) to the  ferromagnetic one ($h_z=0.25$, varying $h_x$);
 (lower panel) quench within the paramagnetic case from $h_z=2$ and $h_x=0$ to $h_z=1.75$ and varying $h_x$.
 While in the confining phase the light cone experiences a
drastic non-perturbative change, in the paramagnetic phase it is only perturbatively modified.
 }
 \label{fig:more} 
\end{figure}

As a further confirmation of the above scenario  
it is natural to consider a very large quench to a confining Hamiltonian in such a 
way that $K(k)$ in Eq. (\ref{IniS}) is not small and mesons with non-zero velocities are formed with high probability. 
In Fig. \ref{fig:more} (top) we report the connected correlations corresponding to a quench from the paramagnetic phase 
($h_z=2, h_x=0$) to the  ferromagnetic confining one ($h_z=0.25$, varying $h_x$). 
In this case the light cones are visible without zooming.  
Their velocities always correspond to the maximal speed of the mesons (reported as a straight line). 
A final check for the validity of the overall scenario is that for quenches to the paramagnetic phase in the 
presence of an external longitudinal field there should not be any strong change in the light-cone since there 
is no confinement. 
This is quite apparent in Fig. \ref{fig:more} (bottom) where we report the data for a quench from $h_z=2$ and $h_x=0$
to $h_z=1.75$ and varying $h_x$. It is clear that adding the magnetic field $h_x$ does not alter the qualitative shape 
of the light-cone.

We have also studied the connected correlation function of the transverse component of the spin 
(density in the fermionic language). This correlation function also reflects the change of the light cone due to 
the modified velocity of the mesons. 
%For lack of space 
We report the corresponding density plots in the Supplementary Material \cite{SM}. Furthermore, we also 
examined quenches from and to several other values of the two magnetic fields in the Hamiltonian, and the overall picture for the correlation functions is found to be the one we extracted from the first few examples, so we stress that our conclusions are very  general and not limited to the reported cases.

\begin{figure}[t]
 \includegraphics[width=.23\textwidth]{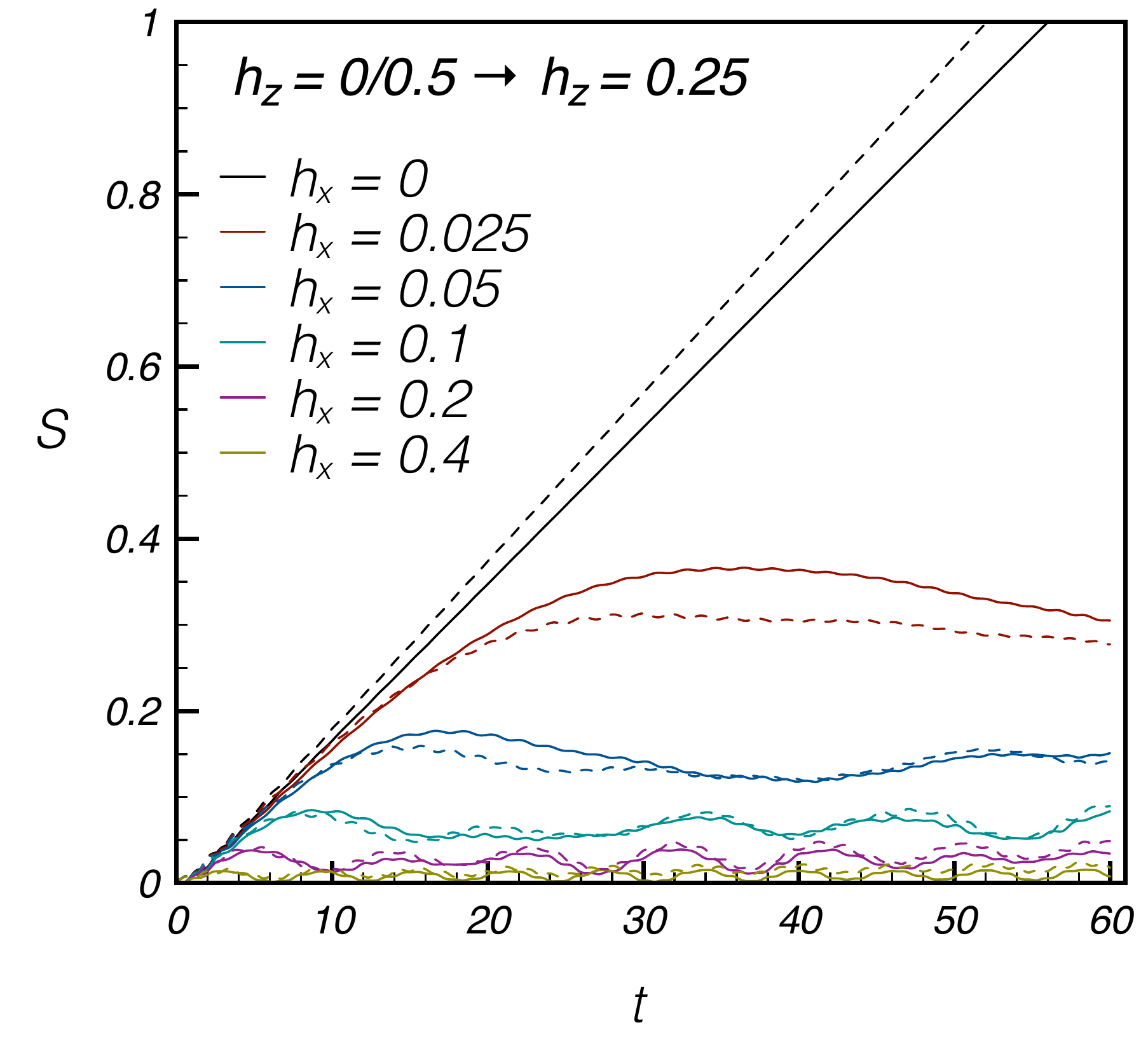}
  \includegraphics[width=.23\textwidth]{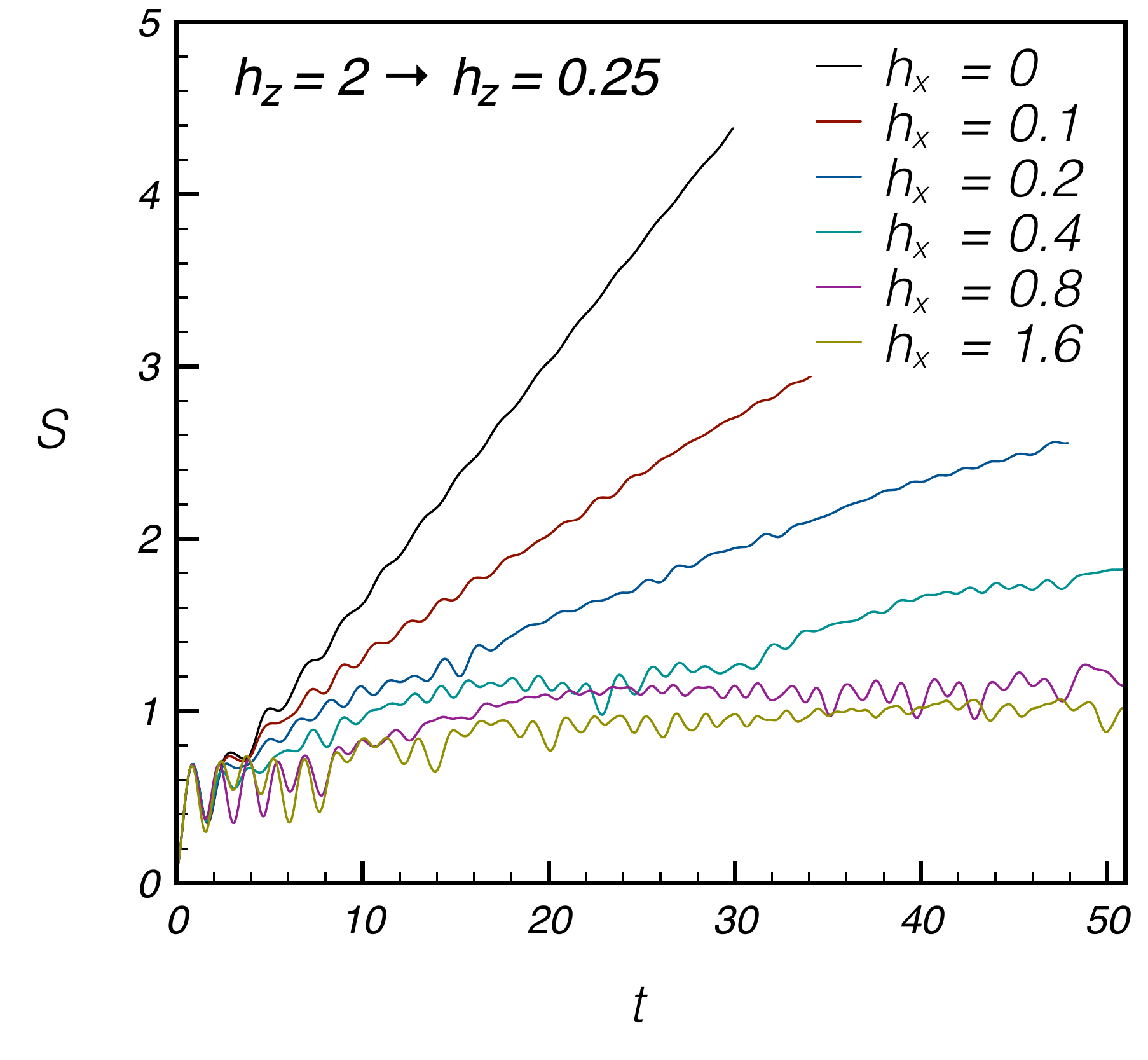}
 \caption{Time evolution of the half-chain entanglement entropy after a quench to the confining phase. 
 Left: starting from the ferromagnetic phase ($h_z=0$).
 Right: starting from the paramagnetic phase $h_z=2$, $h_x=0$. }
\label{fig:EE}
\end{figure}

{\it Entanglement entropy} is another important probe (indeed a true smoking gun) for the quasi-particle propagation and hence light cone effects \cite{cc-05}.
It is defined as the von Neumann entropy $S_A=-{\rm Tr} \rho_A \ln \rho_A$ of the reduced density matrix 
$\rho_A$ of a subsystem $A$.
This can be readily accessed by the iTEBD method, especially for the case when $A$ corresponds to half of the 
system. %, i.e. the so called half-chain entanglement entropy.
The obtained numerical results are reported for three sets of quenches in Fig. \ref{fig:EE},
two within the ferromagnetic phase and one across the critical point to the ferromagnetic phase. 
We consider several different final values of the longitudinal fields.
For zero $h_x$, we observe a pronounced linear growth in time of the entanglement entropy in
perfect agreement with the known exact results \cite{fc-08}.
In all cases, by turning on the interaction $h_z$, the growth of the entanglement entropy is considerably slowed down and practically saturates (during the observation time) for quenches within the ferromagnetic phase. 
The latter correspond to cases in which the light-cone of the two-point function is strongly suppressed (i.e.~practically invisible).
As explained above, this is a consequence of the fact that mesons are predominantly produced at rest and 
then the entanglement just oscillates around a saturation value, as in the left panel of Fig. \ref{fig:EE}. 
Actually the small fraction of mesons with non-negligible velocities should produce a very slow increase of the 
entanglement which however is likely too small to be observed. 
In the case of a quench across the critical point, the increase of the entanglement entropy is only 
reduced because of the production of many mesons with non-vanishing velocities.
Overall, the data for the entanglement are compatible with the confinement scenario drawn for the correlations. 
In \cite{SM} we also report some results for quenches in the non-confining phase to show that 
confinement effects are absent in that case. 

Furthermore, the frequencies of the oscillations of the entanglement entropy are also 
in rough agreement with the meson masses and their differences,
but a more accurate analysis (similar to the case of the one-point function) 
is difficult due to the presence of a long transient and the constant drift. 
In \cite{SM} we show that for a quench only in $h_x$ (i.e. leaving $h_z$ constant), 
these frequencies can be extracted effectively because there is no drift. 
%
%\blue{We have also considered the entanglement entropy of a finite block. 
%The obtained data are compatible with a light-cone with the velocities of the mesons, but the resolution is worse than 
%for the two-point function and for this reason we do not report these results here.} 

{\it Discussions and further developments}.
We have given compelling arguments and numerical evidence showing that confinement  
strongly affects the non-equilibrium dynamics following a quantum quench. 
The main effect is a dramatic change of the light-cone 
structure of correlation functions and entanglement entropy. 
At the same time, the one-point functions oscillates in time 
with frequencies equal to the meson masses. 
These effects should be easily measurable in cold atom experiments: 
we expect that corrections due to the trapping harmonic potential 
should be as small as the almost negligible finite size effects we observed in the numerics.
Furthermore, our results show that the quench dynamics can be used 
(both numerically and experimentally) to probe the confinement and have 
direct access to the meson masses from the power spectrum of the one-point functions.
This `quench spectroscopy' could turn out to be  more powerful than standard equilibrium methods to 
measure the spectrum.

We can  speculate on a few other consequences and applications of our work. 
It was noticed some time ago \cite{bhc-10}, that in some quenches within the  Hamiltonian (\ref{ham})
the system does not  approach asymptotically a thermal stationary state as expected based on the non-integrability of the model. 
One could speculate that because of confinement, there are rare states in the spectrum which prevent
eigenstate thermalisation hypothesis \cite{eth}  to be applied. 
Along the same line of thought, it is also clear that even if these confined systems eventually thermalise, 
the standard prethermalisation scenario \cite{pret} for weak integrability breaking cannot be applied, since 
a  small perturbation not only changes the long time asymptotic expectation values, but
completely alters the dynamics even at short time scales.

Finally, confinement is expected to have similar effects 
also in higher dimensions and so for the theory of strong interactions. 
It is natural to wonder what the consequences are for realistic  non-equilibrium situations in quantum chromodynamics 
such as the quark-gluon plasma in hadron colliders. 
While even approximate field-theoretical calculations for strong interactions are beyond our reach,  
holographic methods have been successfully applied to the study of standard light cone phenomena \cite{hol1}
and to quenches in confining theories \cite{hol2}
could be insightful to understand this fascinating problem.

{\it Acknowledgments}. We are grateful to R. M. Konik and J. Cardy  for helpful discussions.  
This work was supported by the  ERC under  Starting Grant 279391 EDEQS (PC and MC). 
MK thanks SISSA for hospitality and was partially supported by a Janos Bolyai Research
Scholarship of the HAS. This work was also partially supported by the CNR-HAS bilateral grant SNK-84/2013.

%%%%%%%%%%%%%%%%%%%%%%%%%%%%%%%%%%%%%%%%%%%%%%%%%%%%%%
%%%%%%%%%%%%%%%%%%%%%% SUPPLEMENTAL MATERIAL%%%%%%%%%%%%%%%%%
%%%%%%%%%%%%%%%%%%%%%%%%%%%%%%%%%%%%%%%%%%%%%%%%%%%%%%

\def\erf{\eqref}
\newcommand{\expct}[1]{\left\langle #1 \right\rangle}
\newcommand{\vev}[1]{\langle #1 \rangle}
\newcommand{\ud}          {\mathrm d}
\newcommand\eps           {\varepsilon}
\newcommand\w           {\omega}
\newcommand\fii           {\varphi}
\newcommand\mc            {\mathcal}
\newcommand\LL            {Lieb--Liniger }
\newcommand\p             {\partial}
\newcommand\lam		{\lambda}
\newcommand\psid          {\psi^{\dagger}}
\renewcommand\th          {\theta}
\newcommand\kb            {k_\text{B}}
\newcommand \rhop       {\rho^{\text{(r)}}}
\renewcommand{\vec}[1]   {|#1\rangle}
\newcommand{\cev}[1]   {\langle#1|}
\newcommand{\fR}   {f^{\text{R}}}
\newcommand{\fL}   {f^{\text{L}}}
\newcommand{\GR}   {F^{\text{(R)}}}
\newcommand{\GL}   {F^{\text{(L)}}}
\newcommand{\TR}   {T_{\text{R}}}
\newcommand{\TL}   {T_{\text{L}}}
\newcommand{\rhoR}   {\rho_{\text{R}}}
\newcommand{\rhoL}   {\rho_{\text{L}}}
\newcommand{\qR}   {q_{\text{R}}}
\newcommand{\qL}   {q_{\text{L}}}
\newcommand{\NR}   {{N_{\text{R}}}}
\newcommand{\NL}   {{N_{\text{L}}}}
\newcommand{\xR}   {\bar x^{(\text{R})}}
\newcommand{\xL}   {\bar x^{(\text{L})}}
\newcommand{\tR}   {\bar t^{(\text{R})}}
\newcommand{\tL}   {\bar t^{(\text{L})}}
\newcommand{\phiR}   {\phi^\text{R}}
\newcommand{\phiL}   {\phi^\text{L}}
\newcommand{\psiR}   {\psi^\text{R}}
\newcommand{\psiL}   {\psi^\text{L}}
\newcommand{\tphiR}   {\tilde\phi^\text{R}}
\newcommand{\tphiL}   {\tilde\phi^\text{L}}
\newcommand{\tpsiR}   {\tilde\psi^\text{R}}
\newcommand{\tpsiL}   {\tilde\psi^\text{L}}
\newcommand{\sig}  {\sigma}

%%%%%%%%%% Merge with supplemental materials %%%%%%%%%%
\clearpage
\begin{onecolumngrid}
\begin{center}
{\bf\large SUPPLEMENTAL MATERIAL}\\
\vspace{0.1cm}
{\bf\large Real time confinement following a quantum quench to a non-integrable model}\\
\vspace{0.3cm}
M. Kormos,  M. Collura, G. Tak\'acs,  and P. Calabrese
\vspace{1cm}
\end{center}
\end{onecolumngrid}
\begin{twocolumngrid}
%%%%%%%%%% Merge with supplemental materials %%%%%%%%%%
%%%%%%%%%% Prefix a "S" to all equations, figures, tables and reset the counter %%%%%%%%%%
\setcounter{equation}{0}
\setcounter{figure}{0}
\setcounter{table}{0}
\setcounter{page}{1}
\makeatletter
\renewcommand{\theequation}{S\arabic{equation}}
\renewcommand{\thefigure}{S\arabic{figure}}
\renewcommand{\bibnumfmt}[1]{[S#1]}
\renewcommand{\citenumfont}[1]{S#1}

\appendix

\section{Quantum quenches in the transverse field Ising chain}\label{app1}

The Hamiltonian (1) in the main text with $h_x=0$ yields the  integrable transverse field Ising chain which can be diagonalised 
by mapping it (via a Jordan--Wigner transformation) to a system of free spinless fermions  
\be
\label{H_a}
H_\text{TI}=\sum_k \eps(k) a_k^\dag a_k + \text{const.}\,,
\ee
where $a^\dag_k,a_k$ are fermionic creation and annihilation operators and the dispersion relation is given by
\be
\eps(k) = 2J \sqrt{1-2h_z \cos k +h_z^2}\,.
\label{disprel}
\ee
The maximal velocity of these excitations is
\be
v_\text{max} = \max_k\eps'(k)=2J\min(1,h_z)\,.
\ee

The time evolution following a sudden quench of the transverse field $h_z^0\to h_z$ has been analytically solved  
for many observables in \cite{CEF}. 
Here we recall some results from these works that are relevant to our current setup. The initial state is the ground state of the pre-quench Hamiltonian which can be written explicitly in terms of the eigenstates of the post-quench Hamiltonian governing the time evolution. As these eigenstates are free fermion Fock-states, this implies that the initial state can be expanded as a superposition of many-particle states
\be
\label{psi0}
|\psi_0\rangle = \mathcal{N}\prod_{k>0}\left(1+i K(k)a^\dag_{-k}a^\dag_k\right)|0\rangle\,,
%=\mathcal{N}e^{i K(k)a^\dag_{-k}a^\dag_k}|0\rangle
\ee
where the normalisation constant is 
\be
\mathcal{N}^{-2}=\prod_{k>0}\left(1+|K(k)|^2\right)\,.
\ee
The state \erf{psi0} is a coherent superposition of particle pairs created with 
amplitude
\be
K(k) = \tan(\Delta_k/2),
\ee
where $\Delta_k$ is the difference of the Bogoliubov angles in the diagonalisation of the initial and final Hamiltonians,
\be
\cos\Delta_k =
4J^2 \, \frac{h_zh_z^0-(h_z+h_z^0)\cos k+1}{\eps_0(k)\,\eps(k)},
\ee
with $\eps_0(k)$ being the dispersion relation (\ref{disprel}) evaluated at $h_{z}^{0}$.
The density of pairs is given by
\be
n=\sum_{k>0}\frac{|K(k)|^2}{1+|K(k)|^2}\,.
\ee
For small quenches, when $h_z-h_z^0$ is small, the amplitude $K(k)$ and the density of created particles will also be small.
Let us mention two examples which are relevant for the discussion in the main text. 
For $h_z^0=0$ and $h_z=0.25$ the maximum of $K(k)$ is $\max_{k} |K(k)|\sim0.12$, which corresponds to a 
maximum density of modes $\max_k n_k\sim0.014$.
Oppositely for a large quench such as from $h_z^0=2$ to $h_z=0.25$, $K(k)$ diverges for $k$ close to zero.

In Ref. \cite{CEF} the time evolution of the longitudinal magnetisation was calculated analytically. For quenches within the ferromagnetic phase ($h_z^0,h_z<1$) it was found to decay exponentially
\be
\langle\sigma^x(t)\rangle \propto \exp\left[t \int_0^\pi\frac{\ud k}\pi \eps'(k)\log|\cos\Delta_k|\right].
\ee

The equal time correlation function of the magnetisation was also calculated in the so-called space time scaling limit, 
$\ell\to\infty,t\to\infty,$ $v_\text{max}t/\ell$ fixed. For quenches within the ferromagnetic phase it is given by
\begin{multline}
\langle\sigma^x_0(t)\sigma^x_\ell(t)\rangle \simeq\\
 \exp\left[\ell \int_0^\pi\frac{\ud k}\pi \log|\cos\Delta_k|\theta_\text{H}(2\eps'(k) t-\ell)\right]\\
\times\exp\left[2t \int_0^\pi\frac{\ud k}\pi \eps'(k) t\log|\cos\Delta_k|\theta_\text{H}(\ell-2\eps'(k) t)\right].
\end{multline}
Here $\theta_\text{H}$ is the Heaviside $\theta$-function which gives rise to a light cone structure for the connected correlation function visible in the first panel of Fig. 3 of the main text.

%%%%%%%%%%%%%%%%%%%%%%% FIG 1 %%%%%%%%%%%%%%%%%%%%%%% 

\begin{figure*}[t!]
\subfloat[]{\includegraphics[width=.5\textwidth]{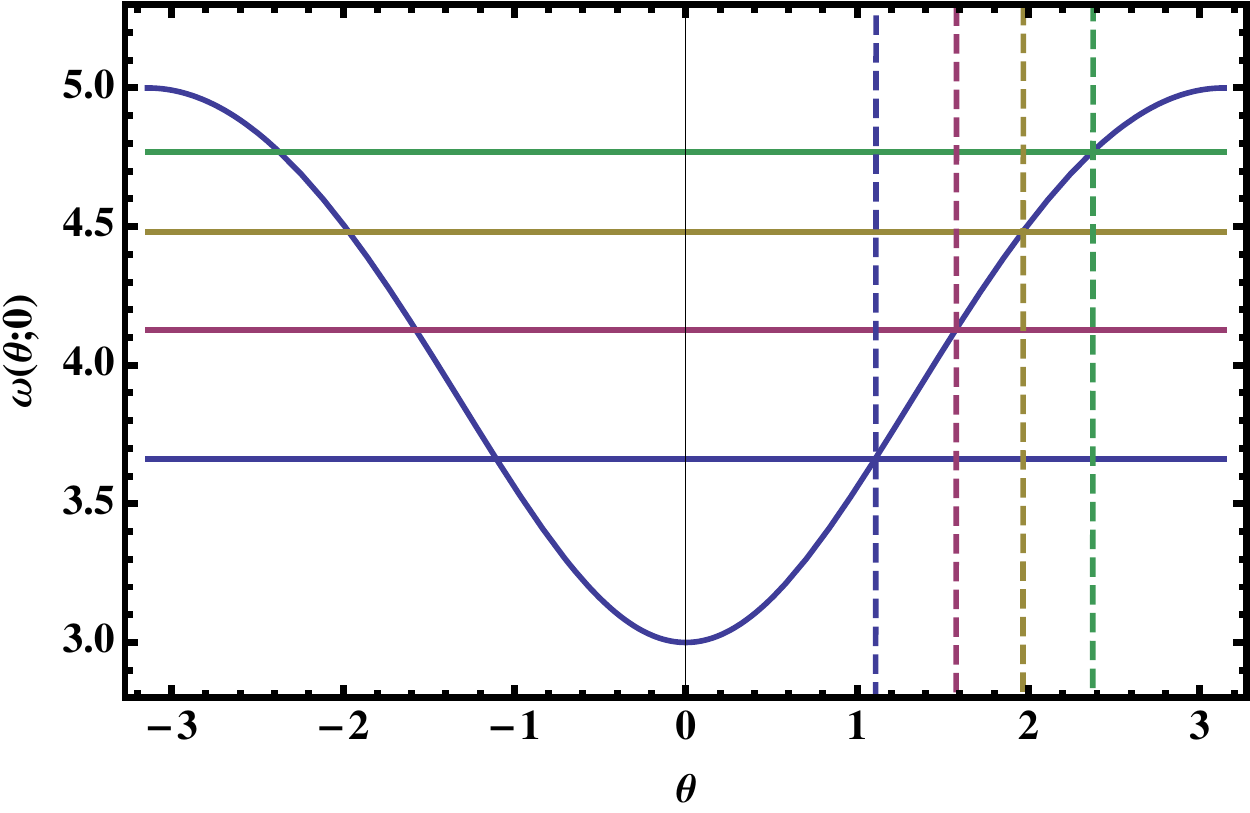}\label{single}}
\subfloat[]{\includegraphics[width=.5\textwidth]{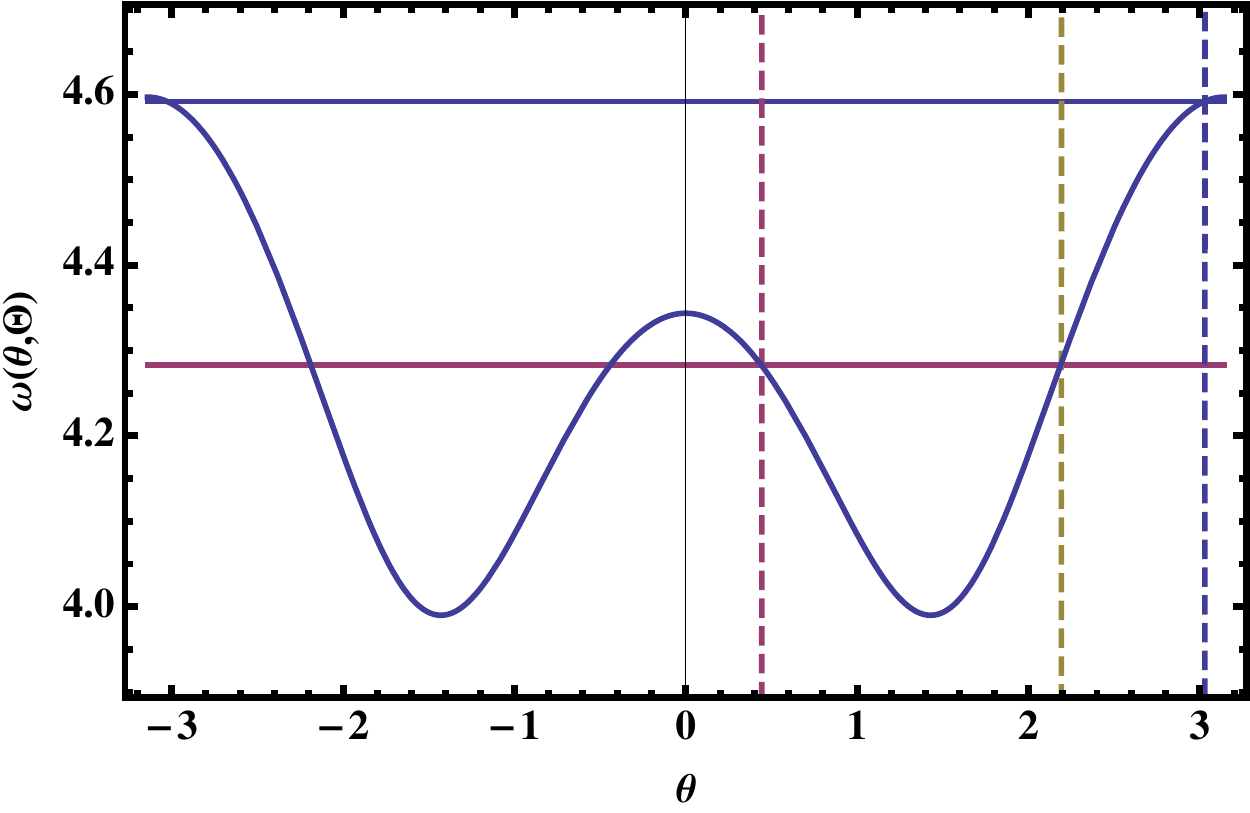}\label{double}}
\caption{Semiclassical bound state energy levels in the ``relative potential'' 
$\w(\th,\Theta)$ from the solutions of Eqs. \erf{singleEqs}.
The dashed vertical lines show the turning points $\th_{a,b}.$ 
(a) Bound states for $h_z=0.25,h_x=0.1,\Theta=0$. 
(b) Bound states for $h_z=0.5,h_x=0.1,\Theta=3$. }
\label{fig:semib}
\end{figure*}
%%%%%%%%%%%%%%%%%%%%%%%%%%%%%%%%%%%%%%%%%%%%%%%%%

%%%%%%%%%%%%%%%%%%%%%%% FIG 2 %%%%%%%%%%%%%%%%%%%%%%%
\begin{figure*}[t!]
\subfloat[][]{\includegraphics[width=.5\textwidth]{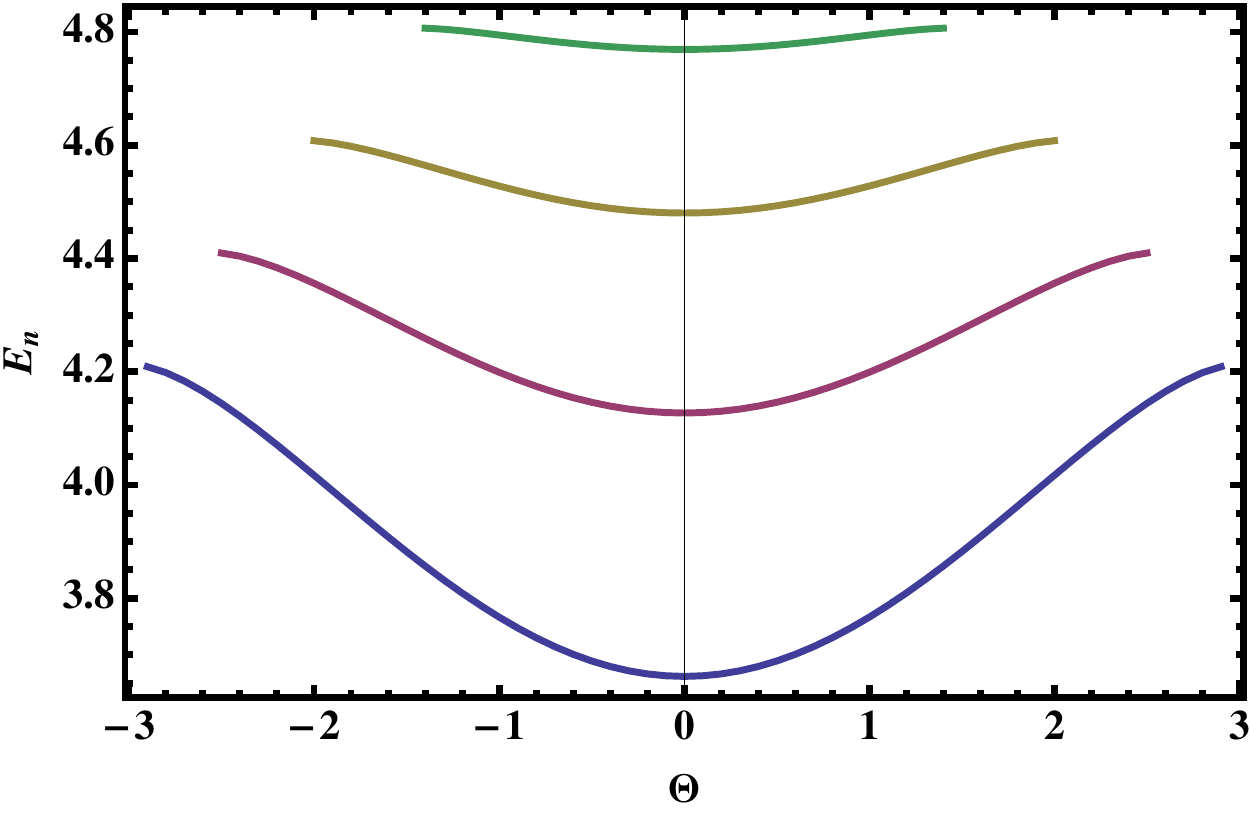}\label{fig:disprel}}\subfloat[][]{\includegraphics[width=.5\textwidth]{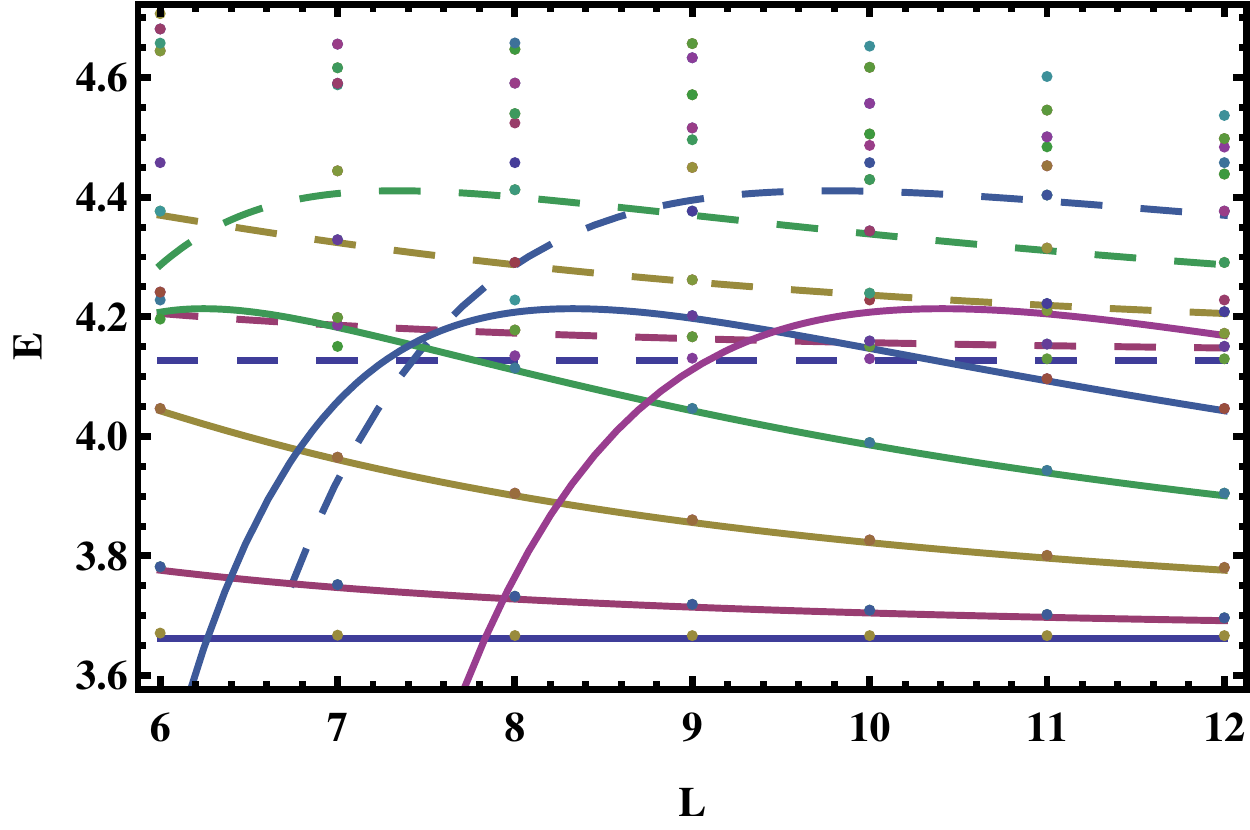}\label{fig:spectr}}
\caption{Semiclassical bound state dispersion relation and the low energy spectrum.
(a) Meson dispersion relations for $h_z=0.25,h_x=0.1$.
(b) Finite size spectrum for $h_z=0.5,h_x=0.1$. Dots are exact diagonalisation results, 
lines are the semiclassical approximation. 
Continuous lines: 1-meson states of the first meson. 
Dashed lines: 1-meson states of the second meson.}
\label{fig:disprel-spectr}
\end{figure*}
%%%%%%%%%%%%%%%%%%%%%%%%%%%%%%%%%%%%%%%%%%%%%%%%%

\section{Semiclassical calculation of the meson dispersion relations}\label{app2}

In the ferromagnetic phase ($h_z<1$) of the transverse field Ising chain, in the absence of longitudinal magnetic field, the elementary excitations are domain walls separating domains of degenerate ground states having opposite non-zero longitudinal magnetisation $\vev{\sigma^x}.$ These domain walls propagate freely in the system with dispersion relation given by Eq. \erf{disprel}. 

Turning on a longitudinal magnetic field causes a non-perturbative change in the particle spectrum, first described by McCoy and Wu \cite{mccoy-wu} in the continuum scaling limit of the model. The main features of this change can be understood based on the following semiclassical argument.
A small non-zero field $h_x$ lifts the degeneracy of the two ferromagnetic ground states, and in particular, a domain with magnetisation opposite to the external field will have energy proportional to its length. Clearly, isolated domain walls have very high energy so they cannot propagate freely anymore but they get confined into bound states. In the literature these bound states are often called  ``mesons'' in analogy to the bound states of strong interactions. 

The number of stable mesons and their dispersion relations can also be approximately computed based on this picture. In the continuum field theory this problem was studied in Refs. \cite{fonseca, rutkevich_continuum}, while in the lattice system it was considered in Refs. \cite{rutkevich1,rutkevich} which we follow closely below.

%%%%%%%%%%% FIGURE DENSITY PLOT SzSz main quench %%%%%%%%%%%%%%%%%%
\begin{figure*}[t!]
\includegraphics[width=1.115\textwidth]{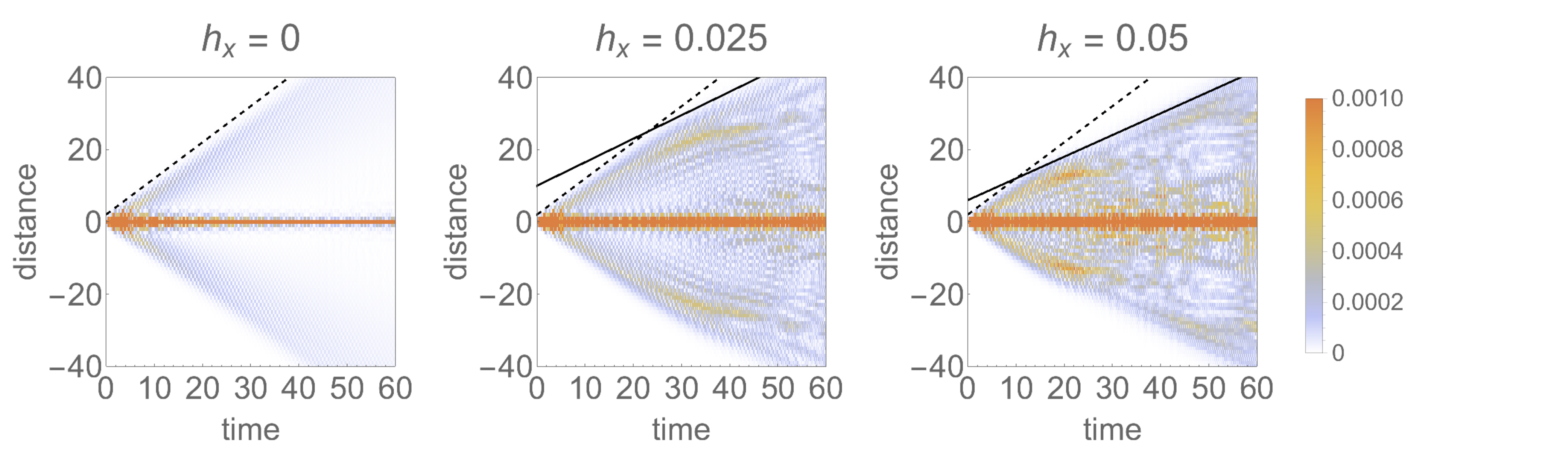}\\
\includegraphics[width=1.115\textwidth]{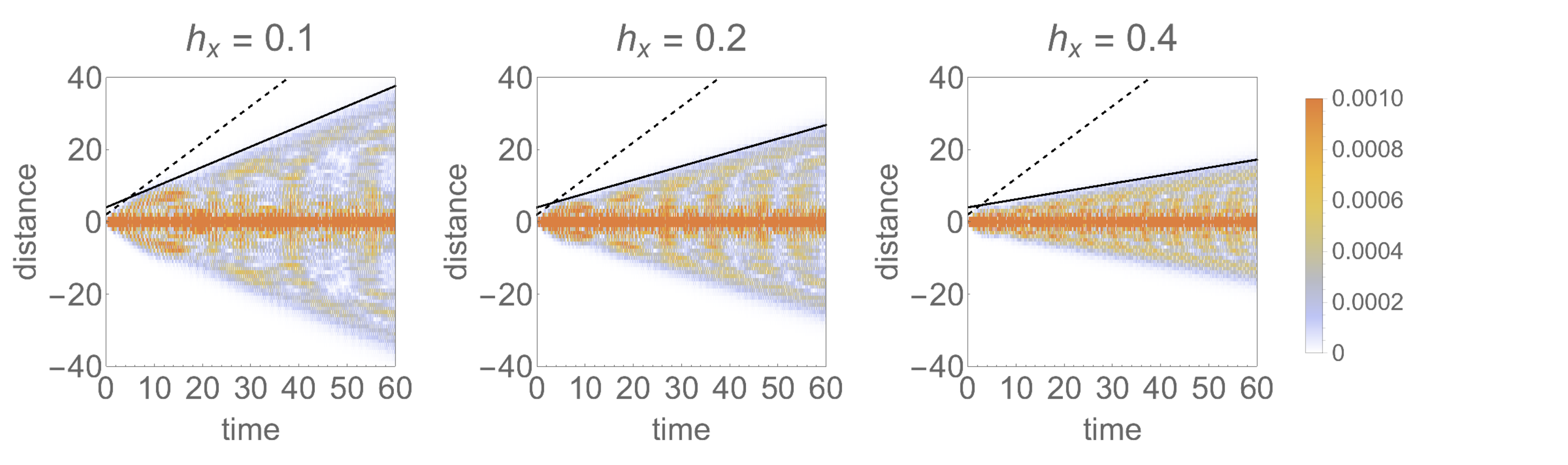}
\caption{Connected transverse spin-spin correlation function 
$\langle \sigma^{z}_{1} \sigma^{z}_{m+1} \rangle_c$ after quenching 
from the maximally ferromagnetic ground state at $h_z=h_x=0$ to the ferromagnetic point $h_{z}=0.25$ 
with different longitudinal magnetic fields $h_{x} = 0,\,0.025,\,0.05, 0.1,\, 0.2,\, 0.4$.}
\label{fig:densitySzSz_main}
\end{figure*}
%%%%%%%%%%%%%%%%%%%%%%%%%%%%%%%%%%%%%%%%%%%%%%%%%%%%

The external field induces a linear attractive potential between neighbouring domain walls which border a domain having magnetisation in the direction opposite to $h_x.$ If $d$ is the distance between the domain walls, the potential is $V(d)=\chi\cdot d$ with $\chi=2Jh_x\bar\sigma.$ Let us now consider two fermions moving in one dimension as a classical system with the Hamiltonian
\be
\mc{H} = \eps(\theta_1)+\eps(\theta_2)+\chi |x_2-x_1|\,.
\ee
For simplicity, the coordinates are taken to be real numbers as in a continuum system, but the dispersion relation is taken to be that in the lattice system. $\theta_1,\theta_2$ are the canonical conjugate variables. After making the canonical transformation
\begin{alignat}{2}
X&=\frac{x_1+x_2}2\,,\qquad& x &= x_2-x_1\,,\\
\Theta&=\theta_1+\theta_2\,,\qquad&\theta &= \frac{\theta_2-\theta_1}2\,,
\end{alignat}
the Hamiltonian takes the form
\be
\mc{H} = \w(\theta;\Theta)+\chi|x|\,,
\ee
where
$\w(\theta;\Theta) = \eps(\theta+\Theta/2)+\eps(\theta-\Theta/2).
$
The canonical equations of motion are
\begin{alignat}{2}
\dot X(t) &= \frac{\p \w(\theta;\Theta)}{\p \Theta}\,,\qquad& \Theta(t)&=\Theta=\text{const.}\,,\\
\dot x(t) &= \frac{\p \w(\theta;\Theta)}{\p \theta}\,,\qquad& \dot \theta(t) &= -\chi\, \mathrm{sgn}(x(t)) \,.\label{eq3}
\end{alignat}

%%%%%%%%%%% FIGURE DENSITY PLOT SzSz other quenches %%%%%%%%%%%%%%%%%%
\begin{figure*}[t!]
\includegraphics[width=1.08\textwidth]{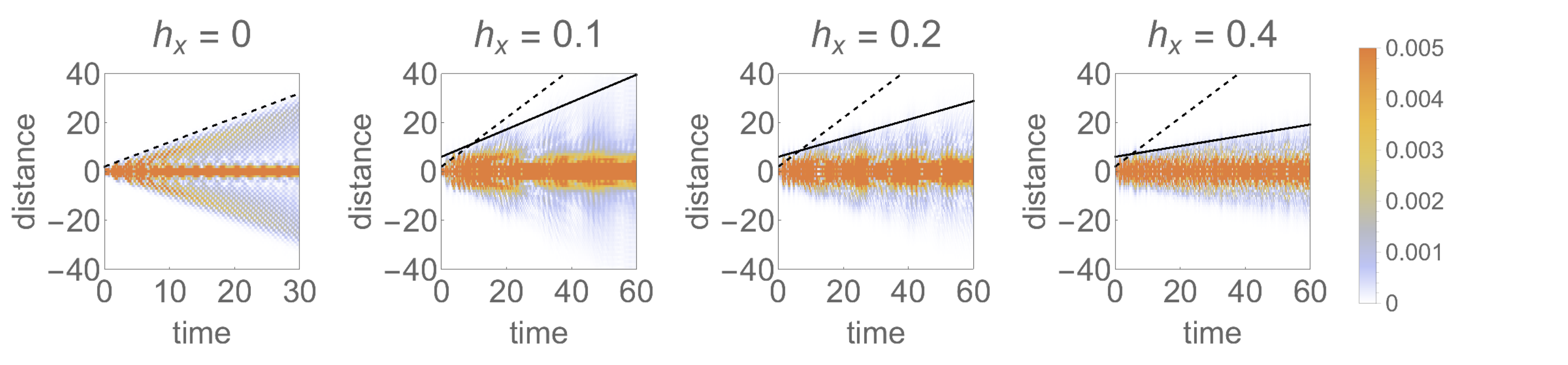}\\
\includegraphics[width=1.08\textwidth]{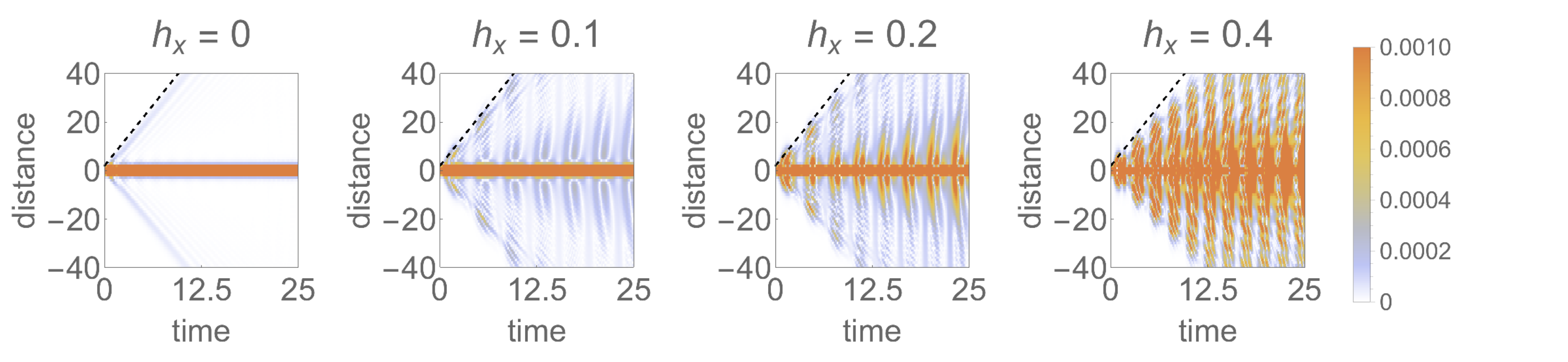}
\caption{Connected transverse spin-spin correlation function 
$\langle \sigma^{z}_{1} \sigma^{z}_{m+1} \rangle_c$ after quenching 
from the paramagnetic point $h_{z}=2$ to the ferromagnetic point $h_{z}=0.25$ (top) and
to the paramagnetic point $h_{z}=1.75$ (bottom) with different longitudinal magnetic fields 
$h_{x} = 0,\, 0.1,\, 0.2,\, 0.4$}
\label{fig:densitySzSz_other}
\end{figure*}
%%%%%%%%%%%%%%%%%%%%%%%%%%%%%%%%%%%%%%%%%%%%%%%%%%%%

For a given value of the total momentum $\Theta$, these equations describe the relative motion of two particles. The solution becomes simple if we think of $q=\theta$ as a spatial coordinate and consider $p=-x$ as the conjugate momentum: we have the periodic motion of a particle with ``kinetic energy'' $\chi|p|$ in the 1D potential $\w(q;\Theta).$ The energy levels can be obtained by the Bohr--Sommerfeld quantization condition which becomes more and more accurate as one moves to higher excited states. When $\Theta< 2\arccos h_z$ the function $\w(q;\Theta)$ has only one minimum at $q=0$ (see Fig. \ref{single}) and this leads to \cite{rutkevich}
\begin{subequations}
\label{singleEqs}
\be
2E_n(\Theta)\th_a -  \int_{-\th_a}^{\th_a}\ud \th\, \w(\th;\Theta) = 2\pi\chi(n-1/4)\,,
\ee
where $n=1,2,\dots$ and the turning point $\th_a=\th_a(n;\Theta)$ is the solution of the equation
\be
\w(\th_a(n;\Theta);\Theta) = E_n(\Theta)\,.
\ee
\end{subequations}
For $\Theta>2\arccos h_z$ the potential $\w(\th;\Theta)$ has two minima (see Fig. \ref{double}).  Then for $E>\w(0;\Theta)$ the above treatment is unchanged, but if $E<\w(0;\Theta),$ the classical motion takes place in one of the two separated wells, and the semiclassical energy levels are given by \cite{rutkevich}
\begin{subequations}
\label{doubleEqs}
\be
E_n(\Theta)(\th_a-\th_b) -  \int_{-\th_b}^{\th_a}\ud \th\, \w(\th;\Theta) = \pi\chi(n-1/2)\,,
\ee
with $n=1,2,\dots$ and
\be
\w(\th_{a,b}(n;\Theta);\Theta) = E_n(\Theta)\,.
\ee
\end{subequations}

The solutions of Eqs. (\ref{singleEqs},\ref{doubleEqs}) give the dispersion relations $E_n(\Theta)$ of the bound states. As an example, we plot the dispersion relations of the four mesons for $h_z=0.25,h_x=0.1$ in Fig. \ref{fig:disprel}. 
The energy gaps (masses in particle physics languages) of these states are 
\begin{eqnarray}
&&m_1=3.662J,\quad m_2=4.127J,\quad\\
&& m_3=4.48J,\quad m_4=4.769J\,.
\end{eqnarray}
They cannot have arbitrarily large momenta $\Theta$, at least semiclassically, and higher lying mesons have flatter dispersion relations. Since their velocities are given by
\be
v_n(\Theta) = \frac{\ud E_n(\Theta)}{\ud \Theta}\,,
\ee
this means that the heavier mesons move more slowly. The maximal velocities of the four mesons are 
\begin{eqnarray}
&&v_1=0.274J,\quad v_2 =0.166J,\\
&&v_3 =0.0937J,\quad v_4=0.0396J,
\end{eqnarray}
thus even the lightest meson has much lower velocity than the 
%For $h_x=0.2$ the velocities are $v=0.188,0.052.$ 
unbounded domain walls which have maximal velocity $v=0.5J$ at this value of the transverse field.

For $h_x=0.2$ the semicalssical approximation provides two mesons. 
Their mass gaps and maximal velocities are
\begin{eqnarray}
&& m_1 = 4.025J,\quad m_2=4.702J,\\
&& v_1 = 0.188J, \quad v_2=0.0518J\,.
\end{eqnarray}

To test the accuracy of this semiclassical analysis, we compare the energies of the mesons with the exact finite size spectrum obtained via exact diagonalisation. In Fig. \ref{fig:spectr} we plot the low energy part of the spectrum as a function of the length of the chain $L$ together with one-particle dispersions $E_{1,2}(2\pi k/L)$ where $k$ is an integer. The agreement is very good even for small lattices and for the lightest mesons.
%It is obvious that the spectrum can be interpreted as energy levels of traveling particles, and also that the semiclassical approximation is quite accurate already for the lightest meson.
%Note that the lightest two-body states lie beyond the plotted energy range having energies $E>2E_1(0,0)\approx7.3.$

\section{Details of the iTEBD calculations}\label{app3}
The iTEBD algorithm \cite{v07}  is based on the infinite Matrix Product State (iMPS) 
description of one dimensional translational invariant lattice 
models in the thermodynamic limit (thus it is free of any finite size effect). 
The canonical iMPS representation of a generic many-body state is 
\begin{equation}
|\Psi\rangle = 
\sum_{\{s\}} 
{\rm Tr}[\cdots {\bf\Gamma}_{o}^{s_{j}} {\bf\Lambda}_{o} 
{\bf \Gamma}_{e}^{s_{j+1}} {\bf\Lambda}_{e} \!\cdots ]
|\cdots s_{j}s_{j+1}\!\cdots\rangle,
\end{equation}
where ${\bf \Gamma}_{o/e}^{s_{j}}$ are $\chi\times\chi$ matrices associated with odd/even lattice sites,
with $s_{j}$ spanning the $j^{\rm th}$-site Hilbert  space in the canonical basis 
$\{|\!\!\uparrow_z\rangle, |\!\!\downarrow_z\rangle\}$; similarly, ${\bf \Lambda}_{o/e}$ are diagonal matrices 
with entries equal to the singular values associated with the bipartition of the system onto the odd/even bonds.

Starting from a given state in MPS representation, the time evolution is obtained with 
 $2^{\text{nd}}$ order Suzuki--Trotter approximation of the evolution operator, namely
\begin{equation}\label{eq:trotter2}
e^{-i H dt } \simeq \bigotimes_{j\,odd}e^{-i \mathfrak{h}  dt/2 }
\bigotimes_{j\, even}e^{-i \mathfrak{h} dt }
\bigotimes_{j\,odd}e^{-i  \mathfrak{h}  dt/2}\,,
\end{equation}
where $\mathfrak{h} $ is the local interaction between nearest neighbour spins. 
Notice that even when the initial state and the post-quench Hamiltonian are one-site shift invariant, 
this is partially broken by the Suzuki--Trotter approximation.
We fixed the Trotter time step at
$dt = 0.005$  (we verified that the data are not affected by the time discretisation). 
As the entanglement increases with time, the auxiliary dimension $\chi$ 
is dynamically updated in order to optimally control the truncation error. 
At each local step, all the Schmidt vectors corresponding to singular values 
larger than $\lambda_{\rm min} = 10^{-16}$ are retained. 
This condition is relaxed when $\chi$ reaches the maximal value $\chi_{\rm max} \in [512,1024]$,
depending on the particular simulation. Because of the upper bound $\chi_{\rm max}$,
the truncation procedure is the main source of error of the algorithm and puts limitations on the 
maximum time accessible by the simulation.

The iMPS representation of initial states has been obtained 
using iTEBD algorithm in imaginary time (except for the particular  states $|\cdots \uparrow_z \cdots\rangle$ 
and $|\cdots \uparrow_x \cdots\rangle$ admitting an exact iMPS representation 
with $\chi_0 = 1$). 
Starting from a simple product state, we evolve it using (\ref{eq:trotter2}) with imaginary 
time $dt = -i \tau = -0.001i$. We verified that decreasing further $\tau$ 
does not affect the final result within our numerical accuracy. 
Since during imaginary time evolution the iMPS loses its canonical form, we implemented 
a procedure to restore it as in \cite{ov08}.
Finally, we checked the convergence of the algorithm by looking at the energy density
and increasing the auxiliary dimension up to $\chi_0 \in [8,16]$. 
The fact that we obtained a very accurate description of the initial state
with a very small auxiliary dimension is mainly due to the mass gap of the considered Hamiltonian.
Indeed, for $h_z=0.25,\, 0.5,\, 2$ the system is far away from the critical point.

%%%%%%%%%%% FIGURE ENTROPIES %%%%%%%%%%%%%%%%%%
\begin{figure}[t!]
\includegraphics[width=0.23\textwidth]{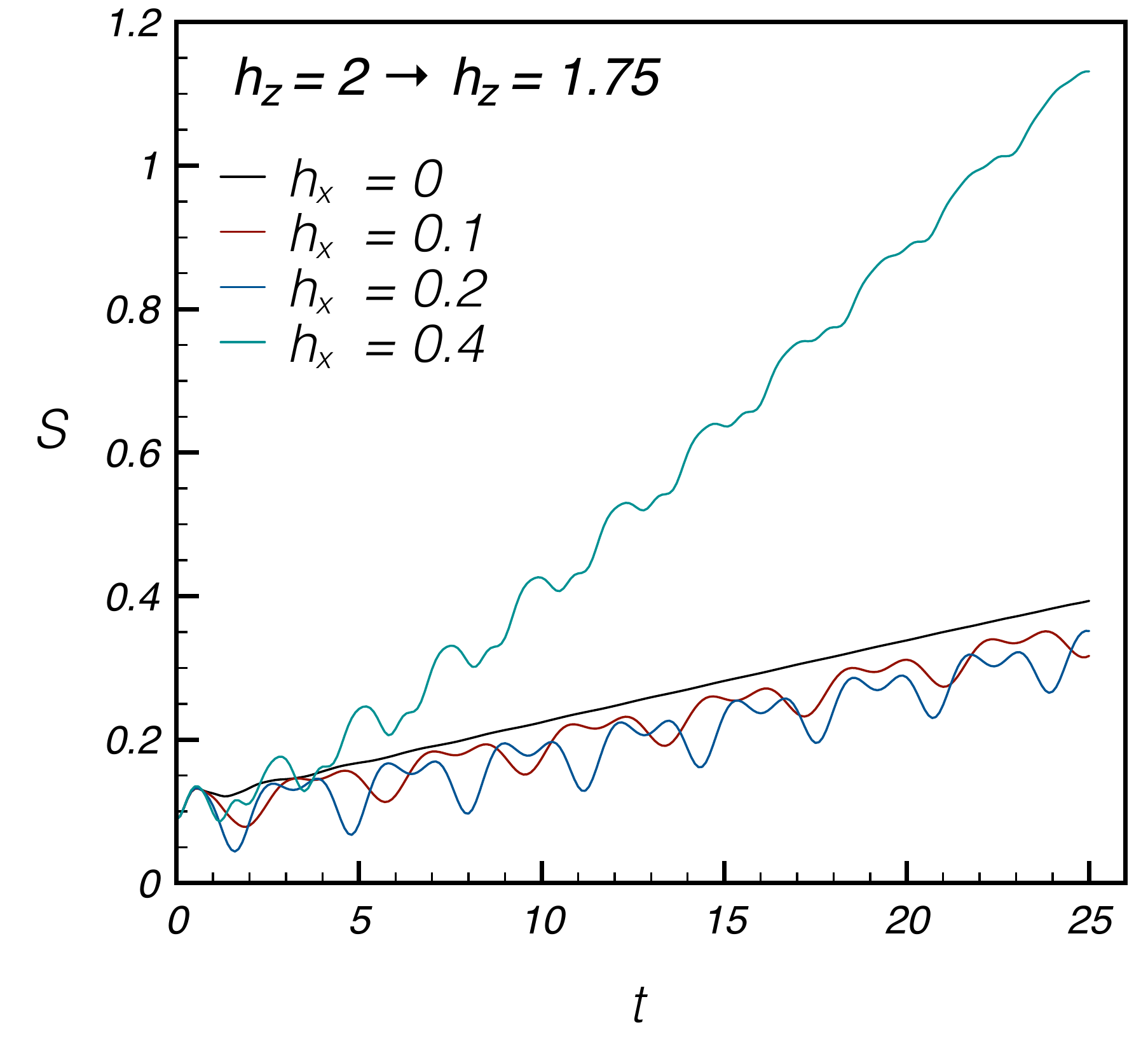}
\includegraphics[width=0.23\textwidth]{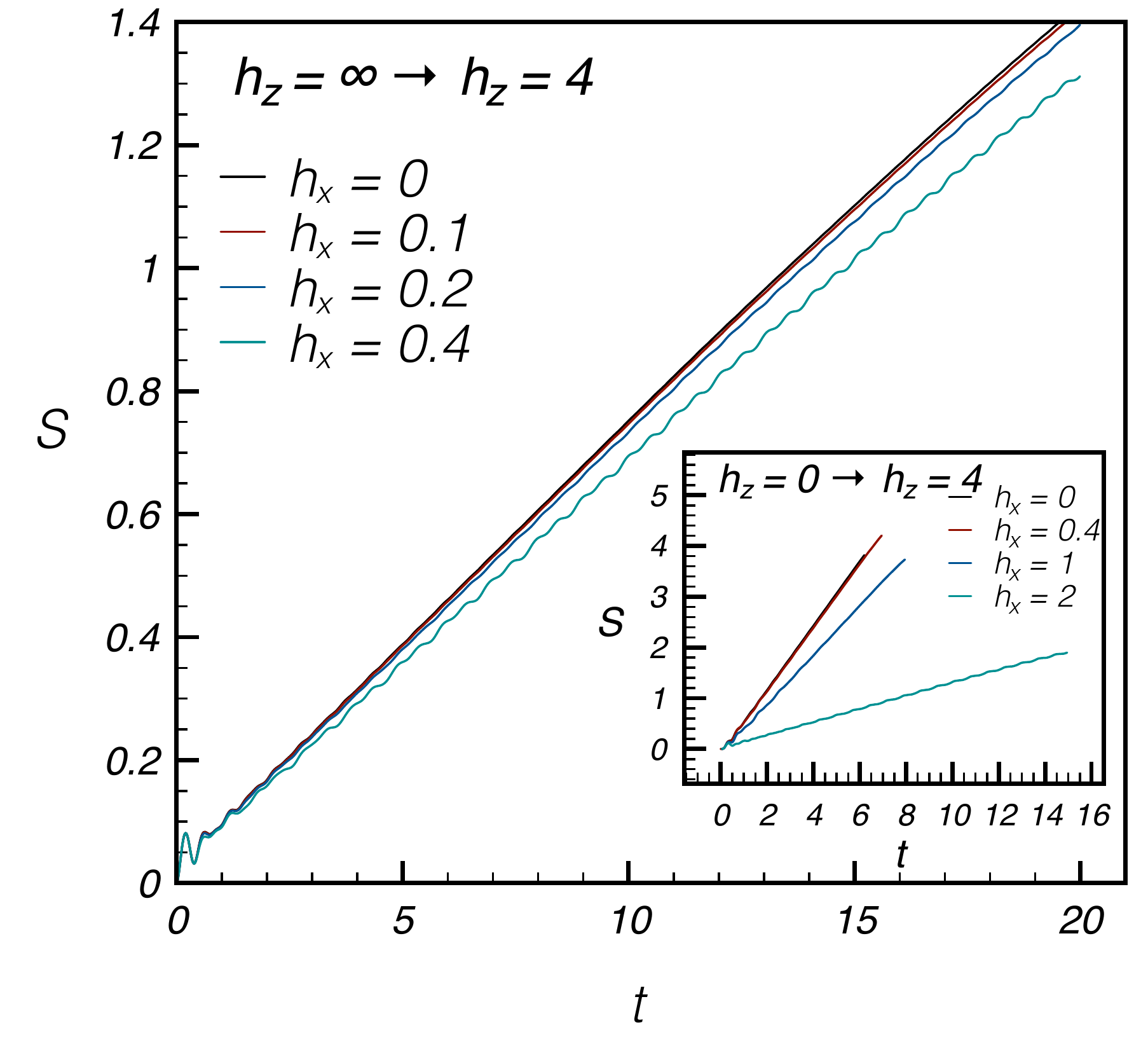}
\caption{Time evolution of the half-chain entanglement entropy 
after a quench within the paramagnetic phase (main figures) and from
the ferromagnetic to the paramagnetic phase (subfigure), for different values of the 
longitudinal field $h_x$.
}
\label{fig:entropy_para}
\end{figure}

\section{Some additional results}\label{app4}

In this section we report some additional results for quantum quenches 
within the Ising Hamiltonian that did not fit in the main text for lack of space.

We start by considering the connected transverse spin-spin correlation function 
$\langle \sigma^{z}_{1}\sigma^{z}_{m+1}\rangle_c$
for a quench within the ferromagnetic phase, 
namely from the initial state fully polarised along $\hat x$ 
(i.e. the ground state at $h_z=0$) to  $h_z=0.25$
and with different values of the longitudinal field $h_x$ (see Fig. \ref{fig:densitySzSz_main}).
As well known from the non-interacting case \cite{CEF},
the overall amplitude of the transverse correlations is smaller than that of the longitudinal ones
(notice a different color scale with respect to Fig. 3 in the main text).
%This is essentially due to the ferromagnetic initial state 
%which is delta-correlated in the transverse direction, 
%i.e. $\langle \sigma^{z}_{1}\sigma^{z}_{m+1}\rangle|_{t=0} = \delta_{m0}$
%(on the contrary $\langle \sigma^{x}_{1}\sigma^{x}_{m+1}\rangle|_{t=0} =1$).
%As a consequence, in the noninteracting case ($h_x=0$), 
%he connected transverse correlation function 
%s inclined to decay very fast, with some noticeable effects 
%only very close to the free-propagating front.
As a confirmation of the confinement scenario, we found 
that the light cone slopes are equal to twice the maximum velocity of the mesons (full lines in the figure)
which is smaller than the domain wall speed (dashed lines in the figure).
In passing we mention that turning on the interaction $h_x$ creates a richer structure inside the light cone
of this correlation (notice the larger signal compared to the non-interacting case).

\begin{figure}[t!]
\includegraphics[width=0.5\textwidth]{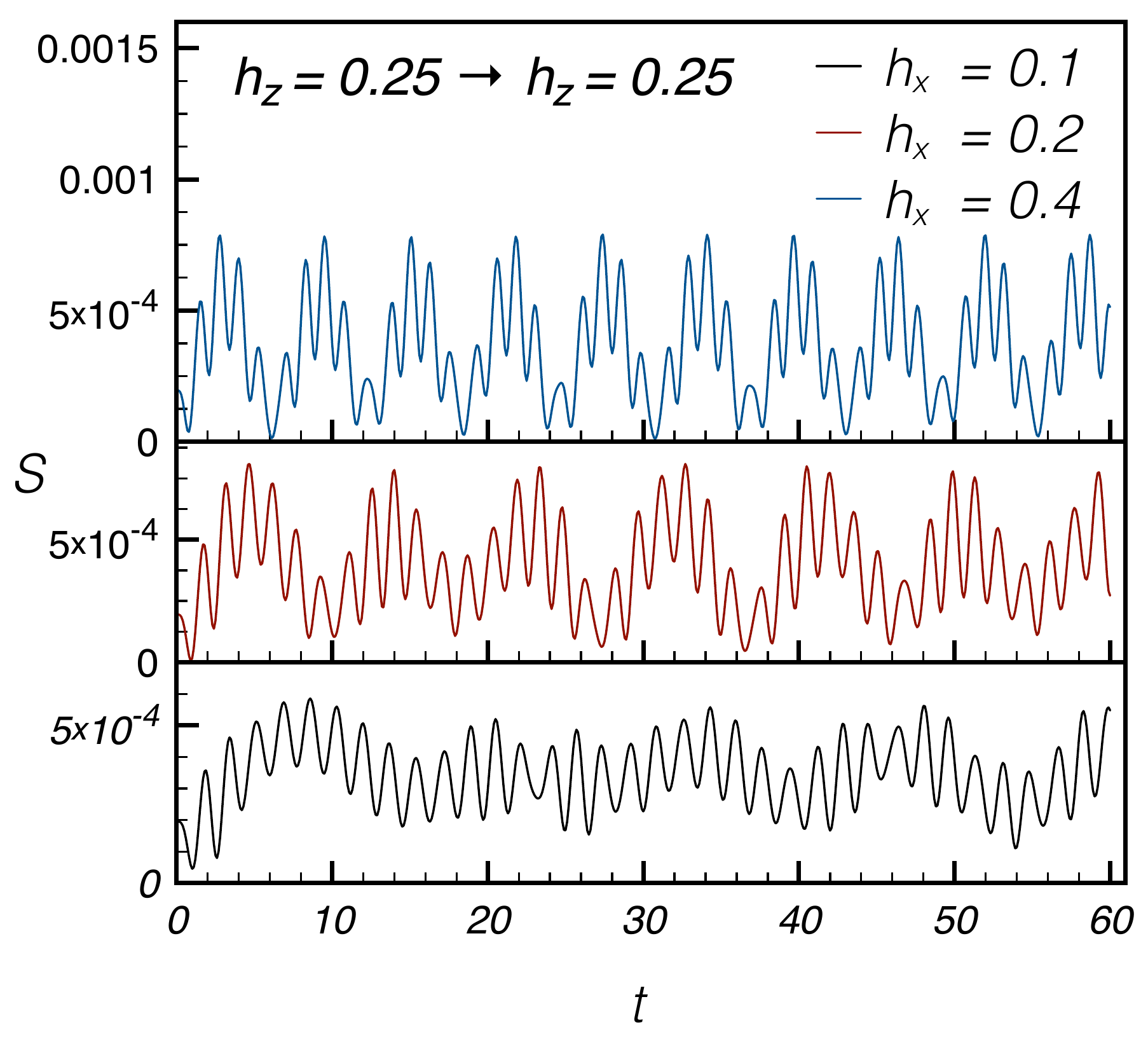}
\caption{The time evolution of the half-chain entanglement entropy after quenching 
only the longitudinal field $h_x$ within the ferromagnetic phase at $h_z=0.25$. The curves 
for $h_x=0.2,\,0.4$ have been vertically shifted for the sake of clarity.}
\label{fig:entropy_time}
\end{figure}
%%%%%%%%%%%%%%%%%%%%%%%%%%%%%%%%%%%%%%%%%%%%%%

%%%%%%%%%%% FIGURE ENTROPIES hz=0.25 SPECTRUM %%%%%%%%%%%%%%
\begin{figure*}[t!]
\includegraphics[width=0.33\textwidth]{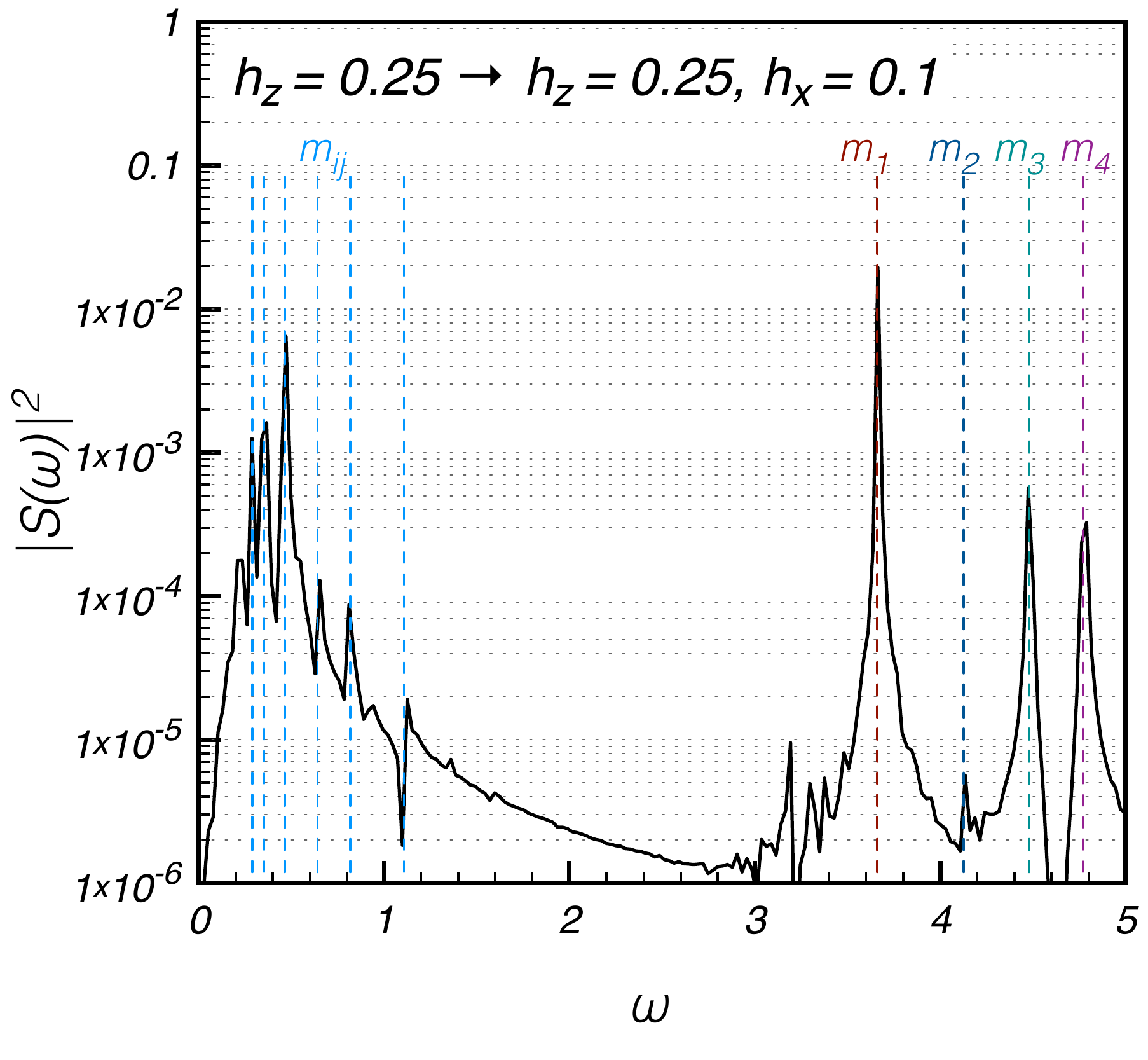}\includegraphics[width=0.33\textwidth]{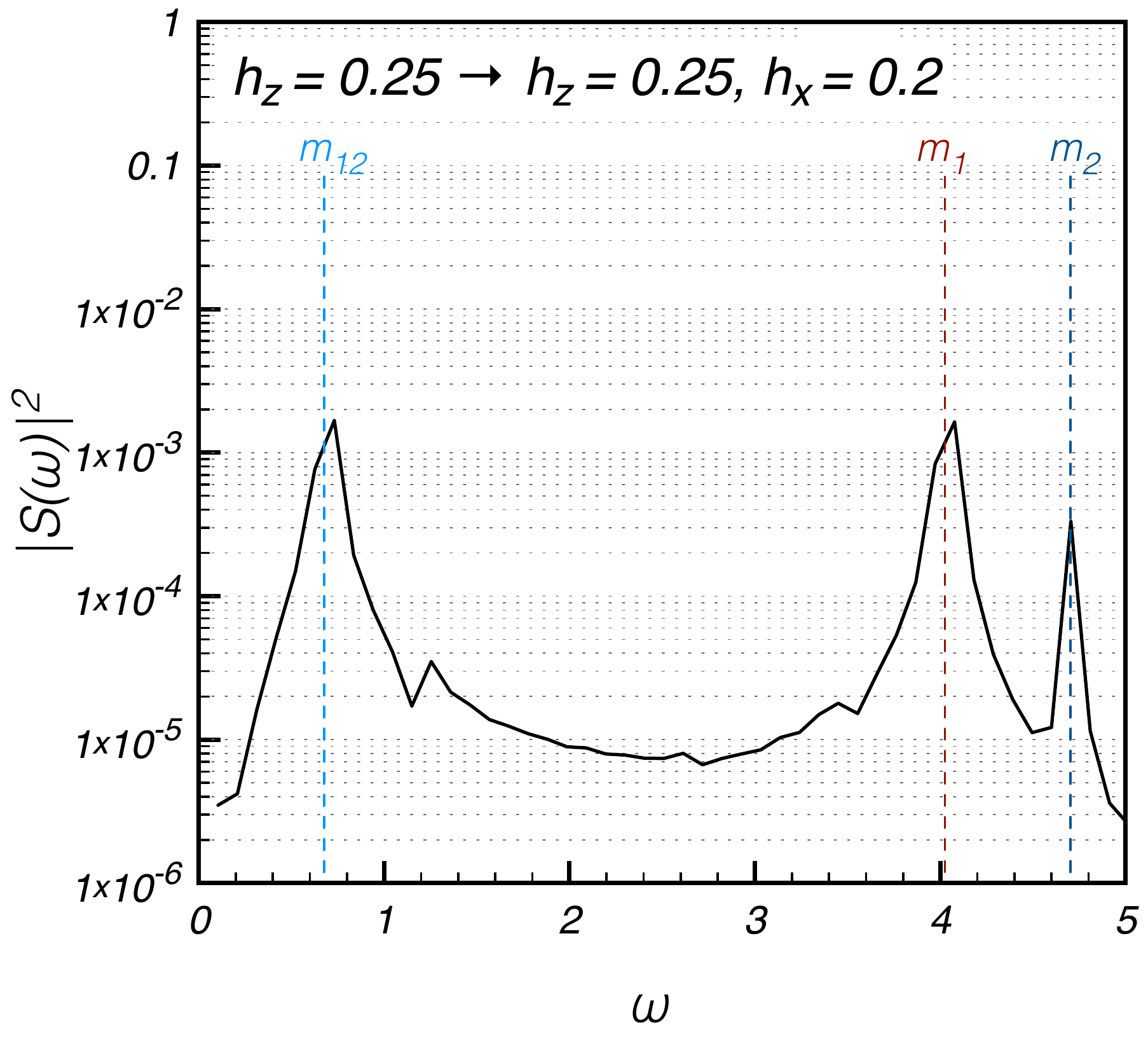}\includegraphics[width=0.33\textwidth]{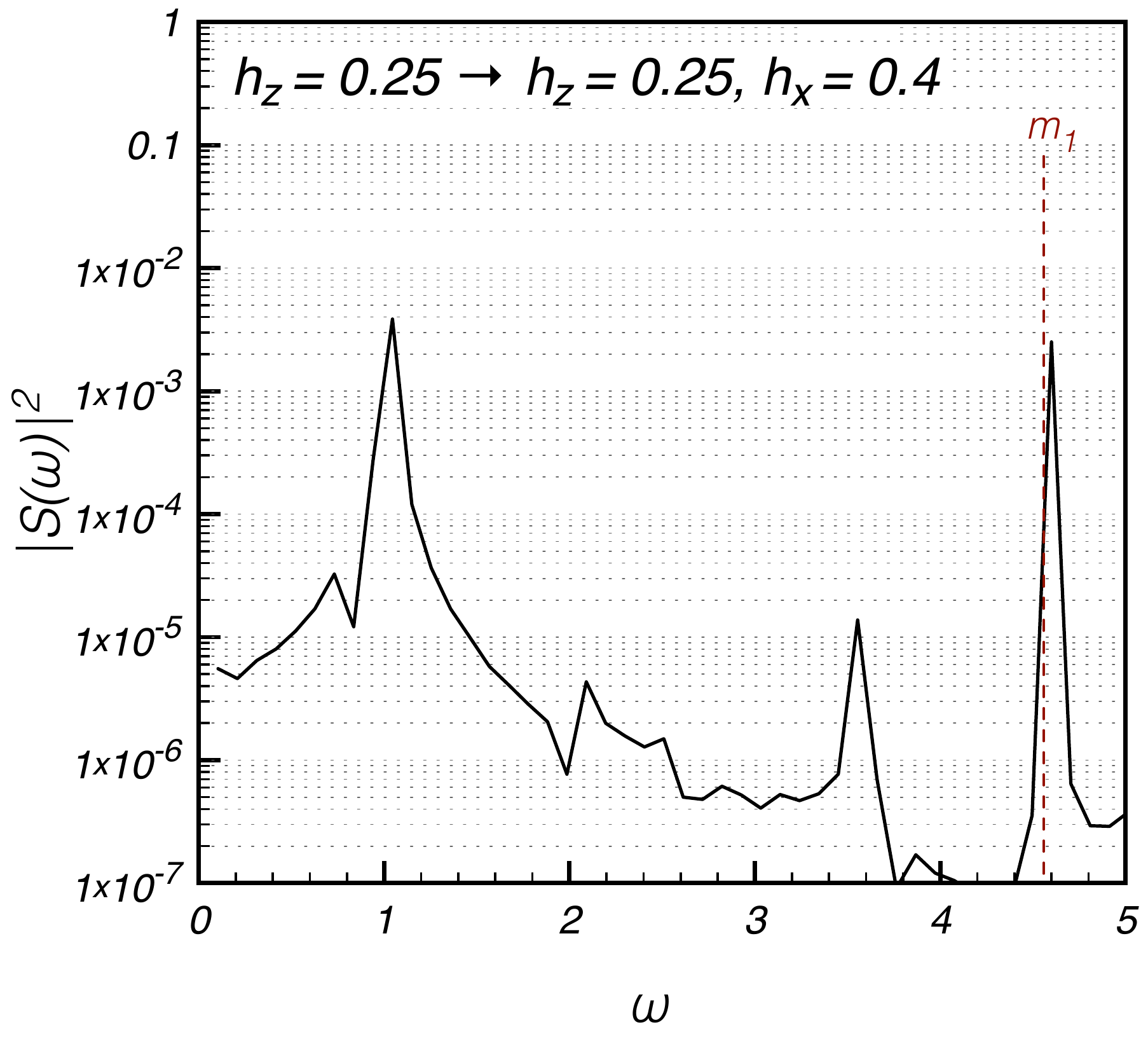}
\caption{Power spectrum of the entanglement entropy after quenching only the longitudinal field
$h_x$. The vertical lines show the meson masses $m_j$ and their differences $m_{ij} = |m_i-m_j|$.
The DFT has been obtained from the time series up to $T=60$ 
for $h_x=0.2,\, 0.4$ and $T=240$ for $h_x=0.1$.}
\label{fig:entropy_spectr}
\end{figure*}
%%%%%%%%%%%%%%%%%%%%%%%%%%%%%%%%%%%%%%%%%%%%%%

Similar considerations apply for all other quenches to the ferromagnetic phase.
For example, in the top of Fig. \ref{fig:densitySzSz_other} we report
the transverse correlations for a quench from the paramagnetic phase ($h_z = 2,\, h_x = 0$) 
to $h_z = 0.25$ and with different values of the magnetic field $h_x$. 
Once again, the slopes of the light cones are compatible with the maximum meson velocities (full lines).  
Moreover, due to the faster spreading of correlations for such a large quench, 
the entanglement entropy grows much faster (see Fig. 6 in the main text) 
causing a breakdown of the iTEBD simulations at earlier times.

Conversely, when the system is quenched to the paramagnetic phase, there is no confinement because of the 
longitudinal magnetic field.  Indeed, the connected transverse correlation functions reported in the 
bottom of Fig. \ref{fig:densitySzSz_other} show that the slope of the light cone is almost unaffected by the 
magnetic field, but the dynamics inside the light cone changes considerably.
Such effects have already been seen in the main text 
for the longitudinal correlation function. 
The absence of confinement after quenching to the paramagnetic phase
is further confirmed by the entanglement entropy. 
In Fig. \ref{fig:entropy_para} we show the time evolution of the 
half-chain entanglement entropy for different quenches to $h_z > 1$. 
Independently from the initial and final Hamiltonians,
the entropy always grows linearly in time and its slope (for small enough $h_x$)
is only slightly perturbed by the presence of a small longitudinal field.

Finally, as anticipated in the main text, we also report some results about the power spectrum of the half-chain 
entanglement for quenches to the ferromagnetic confining phase in which we only quench the longitudinal field $h_x$
keeping $h_z$ fixed at its pre-quench value. 
This choice is motivated by the fact that in these quenches transient and drift effects are reduced, i.e.  
by keeping the transverse field constant we can isolate the oscillating behaviour 
of the domain walls induced by $h_x$. 
In Fig. \ref{fig:entropy_time} we show the time evolution of the half-chain entanglement 
entropy for quenches from $h_z=0.25,\,h_x=0$ to $h_z=0.25,\, h_x=0.1,\,0.2,\,0.4$.
The spectral analysis of the time series of the entanglement entropy $S(t)$ is shown in Fig. \ref{fig:entropy_spectr}.
In particular, for $h_x=0.1$ we used a very long time series (up to $T=240$) for the discrete Fourier transform (DFT) to have a more refined frequency resolution. 
We are then in position to resolve numerically the four meson masses and the six differences.
The agreement with the low density approximation is really impressive. 
For $h_x=0.2$, we can nicely resolve the two masses and their difference. 
Instead for $h_x=0.4$ we notice that we can clearly see the peak corresponding to the mass $m_1$, 
%coming from the treatment explained in Sec. \ref{app2}, 
but we also see two other peaks which signal the presence of 
at least another bound state which is not found  in the semiclassical treatment. 
This does not come completely unexpected because $h_x=0.4$ is a relatively high value of the longitudinal field 
for which the linear approximation for the confining potential is probably not so accurate. 
To conclude, we mention that for all the analysed data sets (for entropy and magnetisation), 
the higher peaks in the power spectra always correspond to the first meson mass $m_1$ and to the first difference 
$m_{12}$ (provided $m_2$ exists).

\end{twocolumngrid}

%%%%%%%%%%%%%%%%%%%%%%%%%%%%%%%%%%%%%%%%%%%%%%%%%%%%%%
%%%%%%%%%%%%%%%%%%%%%%%%%%%%%%%%%%%%%%%%%%%%%%%%%%%%%%
%%%%%%%%%%%%%%%%%%%%%%%%%%%%%%%%%%%%%%%%%%%%%%%%%%%%%%

\end{document}